\documentclass[a4paper,11pt]{article}
\usepackage{jcappub}

\usepackage{orcidlink}
\usepackage{amsmath}
\usepackage{graphicx}
\usepackage[colorlinks=true]{hyperref}
\usepackage{pifont}
\title{Axion Inflation: Perturbative control \\ in the strong backreaction regime}

\author[\diamondsuit]{Valerie Domcke \orcidlink{0000-0002-7208-4464},}
\author[\spadesuit]{Alexandros Papageorgiou \orcidlink{0000-0002-2736-3026},}
\author[\clubsuit]{\\Marco Peloso \orcidlink{0000-0002-9348-9970},}
\author[\heartsuit]{and Stefan Sandner \orcidlink{0000-0002-1802-9018}}

\affiliation[\diamondsuit]{CERN Theoretical Physics Department, 1 Esplanade des Particules, Geneva, Switzerland}
\affiliation[\spadesuit]{Instituto de F\'{i}sica T\'{e}orica UAM-CSIC, c/ Nicol\'{a}s Cabrera 13-15, 28049, Madrid, Spain}
\affiliation[\clubsuit]{Dipartimento di Fisica e Astronomia “G. Galilei”, Universit\`{a} degli Studi di Padova, via Marzolo 8, I-35131 Padova, Italy}
\affiliation[\clubsuit]{INFN, Sezione di Padova, via Marzolo 8, I-35131 Padova, Italy}
\affiliation[\heartsuit]{Theoretical Division, Los Alamos National Laboratory, Los Alamos, NM 87545, USA}

\emailAdd{valerie.domcke@cern.ch}
\emailAdd{papageorgiou.hep@gmail.com}
\emailAdd{marco.peloso@pd.infn.it}
\emailAdd{stefan.sandner@cern.ch}

\abstract{We study the strong backreaction regime in axion inflation, in which the friction from Abelian gauge fields generated via the pseudo-scalar interaction $\frac{\beta}{4 M_p} \phi F {\tilde F}$ is comparable to Hubble friction. The non-linear dynamics associated with the gauge fields render this regime phenomenologically particularly interesting, but also notoriously difficult to study with perturbative methods. Quantifying the perturbative control through the direct impact on the spectral backreaction using (i) a first-order gradient-expansion formalism including axion gradients, and (ii) a one-loop in-in calculation, we discover an extended mild backreaction regime at large couplings $\beta$, with a relatively large, nearly constant particle production parameter $\xi$. In passing, we point out that CMB non-Gaussianity bounds impose a fairly general upper limit on the product of axion gauge field coupling and Hubble parameter during inflation, $\beta H/M_p < 1.4\cdot 10^{-3}$, independently of the choice of the axion potential.
}

\makeatletter
\gdef\@fpheader{%
  \makebox[\textwidth]{%
    Prepared for submission to JCAP%
    \hfill
    CERN-TH-2026-181,%
    \;
    LA-UR-26-26791
  }%
}
\makeatother

\begin{document}
\maketitle
\flushbottom

\section{Introduction}
\label{sec:intro}

Axion-like fields provide a particularly compelling setting in which to study inflationary dynamics beyond the minimal single-field paradigm. Their approximate shift symmetry can protect the flatness of the inflaton potential~\cite{Freese:1990rb}, while permitting derivative and topological interactions with other sectors. A prominent example is the pseudoscalar interaction
$-(\beta/4M_p)\phi F_{\mu\nu}\widetilde F^{\mu\nu}$, which allows the rolling axion $\phi$ to transfer energy efficiently to an Abelian gauge field $F_{\mu\nu}$~\cite{Turner:1987bw,Garretson:1992vt,Anber:2006xt}. For a monotonic axion trajectory, one gauge-field helicity experiences a transient tachyonic instability, controlled by the particle-production parameter
$\xi \equiv \beta\dot\phi/(2M_pH)$. The resulting exponential amplification can source scalar and tensor perturbations~\cite{Barnaby:2010vf}, generating highly non-Gaussian~\cite{Barnaby:2010vf,Barnaby:2011vw} and parity-violating~\cite{Sorbo:2011rz} signals, and, at sufficiently small scales, lead to potentially observable gravitational waves (GW)~\cite{Cook:2011hg,Barnaby:2011qe,Domcke:2016bkh,Bastero-Gil:2022fme,Ozsoy:2024apn} or primordial black holes (PBH)~\cite{Linde:2012bt,Bugaev:2013fya,Garcia-Bellido:2016dkw,Garcia-Bellido:2017aan,Ozsoy:2023ryl,Franciolini:2026cps,Kawasaki:2026gkn}. This mechanism therefore offers a direct link between the microphysics of the inflaton sector and a broad range of cosmological observables.

The same exponential sensitivity that makes gauge production phenomenologically interesting also renders the dynamics intrinsically nonlinear. At sufficiently small $\xi$, the axion follows the usual slow-roll attractor. As $\xi$ grows, however, the gauge fields amplified by the motion of the axion backreact on its evolution through their ${\cal B} \equiv \langle \vec E \cdot \vec B\rangle$ correlator, that acts as an additional friction term in the axion equation of motion~\cite{Anber:2009ua}. Most analytical and semi-analytical studies of this regime adopt the so called {\it homogeneous backreaction} approximation that neglects the axion inhomogeneities, whereas the full spacetime dependence of the gauge field is retained. An interesting aspect of backreaction is that it is not instantaneous. The value of ${\cal B}$ at any given time is sensitive to the growth of the gauge field in the few e-folds that preceded this time, thus introducing a memory effect in the evolution. This delayed response can produce a characteristic sequence of oscillations in the axion velocity and, correspondingly, in the production parameter $\xi$~\cite{Cheng:2015oqa,Notari:2016npn,DallAgata:2019yrr,Domcke:2020zez,Peloso:2022ovc}. These oscillations have been associated with recurring bursts in the gauge field production and the related GW phenomenology~\cite{Garcia-Bellido:2023ser,Barbon:2025wjl}. A notable exception to this picture arises if the gauge field is sufficiently massive, which suppresses this memory effect even in the strong backreaction regime~\cite{Baker:2026xsd,Ferreira:2026klg}).

This oscillatory pattern requires a coherency that can be broken by axion inhomogeneities. The complete dynamics of the system, retaining also the full spacetime dependence of the axion, must be solved via numerical lattice computations~\cite{Caravano:2021bfn,Caravano:2022epk,Figueroa:2023oxc,Caravano:2024xsb,Sharma:2024nfu,Figueroa:2024rkr,Lizarraga:2025aiw,Jamieson:2025ngu,
Kawasaki:2026gkn}. These simulations are extremely challenging, due to the necessity of covering both the relevant infrared (IR) and ultraviolet (UV) scales on a background that is nearly exponentially expanding. As a consequence, also the highly advanced solutions obtained in~\cite{Figueroa:2023oxc,Figueroa:2024rkr} cover ``only'' about $\sim 7$ e-folds of inflation. The requirement of consistently initializing the system and covering the exit from inflation has thus limited the range of couplings that can be explored. In the examples obtained in these works, the gauge fields significantly source axion inhomogeneities, that destroy the oscillatory pattern essentially after the first oscillation.

The gauge field amplification is exponentially sensitive to the parameter $\xi$, and therefore to the axion potential and the axion-gauge coupling. The expensive nature of lattice simulations, and the current limitation in the covered dynamical range, do not allow a full exploration of this highly model-dependent phenomenology. This motivates studies within the much quicker homogeneous backreaction approximation, that however need to be supplemented by some means to assess their accuracy. The present work aims to provide improved diagnostics that can be used for such assessment. 

Possibly the most direct way to assess the importance of axion inhomogeneities is to evaluate the axion gradient energy. Ref.~\cite{Domcke:2023tnn} has shown that homogeneous backreaction can well reproduce the dynamics of the few existing lattice simulations as long as the ratio between the axion gradient and kinetic energy densities remains smaller than a few percent. Ref.~\cite{Barbon:2025wjl} has then shown that, depending on the axion potential and coupling to gauge fields, homogeneous backreaction allows for solutions with several oscillations of the axion velocity that respect this bound. The threshold set for triggering this criterion is however a heuristic one, based on a comparison performed only over a few examples. Despite the gradient energy is an obviously good indicator of the relevance of gradient terms, in this work we concentrate on something more directly related to the axion dynamics, namely on the backreaction term ${\cal B}$. We organize the perturbative expansion directly in terms of the correction to this correlator. Denoting by $\mathcal B^{(0)}$ the result obtained in the homogeneous-backreaction approximation and by $\delta^{(1)}\mathcal B$ the leading correction generated by axion inhomogeneities, we regard the homogeneous treatment as controlled only while
$|\delta^{(1)}\mathcal B|\ll|\mathcal B^{(0)}|$. Operationally, we use a ten-percent correction as the onset of a relevant nonlinear effect. 

We implement this criterion in two complementary ways. The first implementation uses the gradient expansion formalism (GEF)~\cite{Sobol:2019xls}, which replaces the evolution of individual gauge modes by a tower of equations for gauge-field two-point correlators with increasing number of spatial derivatives, that has shown to provide an efficient description of homogeneous backreaction~\cite{Gorbar:2021rlt,Durrer:2023rhc,vonEckardstein:2023gwk}. We employ the extension of this formalism to first order in axion gradients~\cite{Domcke:2023tnn}, and evaluate $|\delta^{(1)}\mathcal B| / |\mathcal B^{(0)}|$ from the GEF evolution with and without the axion gradient terms. The second implementation is a one-loop in-in computation in which the homogeneous-backreaction solution is taken as the unperturbed background and the interaction between the gauge field and the inhomogeneous axion is treated perturbatively. The two implementations are sensitive to the same physical effect from different perspectives: the GEF computation follows the impact of axion gradients on the coupled evolution, whereas the in-in calculation resolves the leading correction to the backreaction spectrum.

We evaluate the two implementations for quadratic axion potential and for different values of the axion-gauge coupling $\beta$. For values of the coupling for which lattice simulations exist in the literature, the two implementations are in very good agreement with each other and accurately point out the moment at which homogeneous backreaction stops providing an adequate description of the full dynamics. Moreover, they corroborate the threshold on the gradient vs.\ kinetic axion energies pointed out in~\cite{Domcke:2023tnn} and the in-in calculation recovers the redistribution of the gauge spectrum from infrared to higher momenta highlighted in the lattice studies~\cite{Figueroa:2023oxc,Figueroa:2024rkr} as key to the breakdown of the homogeneous backreaction regime.

With the aid of these new diagnostic tools, we find a regime at large couplings with a qualitatively new background behavior, characterized by a prolonged phase of mild backreaction in which $\xi$ remains large (hence, possibly leading to a visible phenomenology) and evolves slowly (without oscillations), while gauge friction is significant but remains subdominant to Hubble friction. In this regime, the in-in evaluation reveals a significant enhancement of the backreaction, however without any sign of the infrared to ultraviolet cascade. Consistently, the gradient expansion evaluation does not detect any significant change in the inflationary trajectory compared to the homogeneous backreaction approximation.

Our exploration of the parameter space is guided by analytic considerations. In the limit of (nearly) constant $H$ and $\xi$, the correction to the spectral backreaction in the in-in implementation can be expressed analytically, leading to an upper bound on the perturbative regime in the $(\xi,\beta H/M_p)$ plane. Under the same assumptions, the strong backreaction regime is found to lie in the region of non-perturbativity for $\xi \gtrsim 2.5$ or, equivalently, $\beta H/M_p \lesssim 1$. We complement these findings with a model independent bound (not relying on any specific inflationary potential or on an adiabatic evolution of $\xi$ and $H$ beyond CMB scales) which imposes $\beta H/M_p\lesssim 1.4\cdot 10^{-3}$  from CMB normalization and the upper bound on gauge-sourced scalar non-Gaussianity~\cite{Barnaby:2010vf,Planck:2015zfm}. Together, this implies that (under the assumptions of nearly constant $\xi$ and $H$) the gauge friction in the strong backreaction regime always receives significant corrections compared to the homogeneous backreaction approximation. The impact on the dynamics can then be evaluated using the extended GEF formalism, the full spectral information from the in-in formalism or lattice calculations.

The remainder of the paper is organized as follows. In Sec.~\ref{sec:axion-inflation-review} we review axion inflation with Abelian gauge fields in the homogeneous-backreaction approximation. In Sec.~\ref{sec:axion-inflation-nonlinear} we introduce the perturbativity criterion and its GEF and in-in implementations. Sec.~\ref{sec:results} presents the numerical comparison with lattice simulations, the exploration of stronger couplings, and the analytic bounds obtained for approximately constant $H$ and $\xi$, together with the CMB constraint on $\beta H/M_p$. We summarize our conclusions in Sec.~\ref{sec:conclusions}, while technical details of the in-in loop computation are collected in App.~\ref{app:loop}.

\section{Axion inflation in the homogeneous backreaction regime}
\label{sec:axion-inflation-review}

We dedicate this section to a brief review of the physics of axion inflation coupled with Abelian gauge fields in order to establish our notation and set the stage for a numerical and analytic exploration of nonlinear effects in the latter sections. Our starting point is the action
\begin{equation}
    S=\int d^4 x \sqrt{-g}\left[\frac{M_p^2}{2}R-\frac{1}{2}\partial_\mu\phi\partial^\mu \phi-V(\phi)-\frac{1}{4}F_{\mu\nu}F^{\mu\nu}-\beta\frac{\phi}{4 M_p}F_{\mu\nu}\tilde{F}^{\mu\nu}\right] \,,
\label{action}
\end{equation}
where $F_{\mu\nu}\equiv\partial_\mu A_\nu-\partial_\nu A_\mu$ is the field strength tensor of the gauge field and its dual $\tilde{F^{\mu\nu}}\equiv \frac{\epsilon^{\mu\nu\alpha\beta}F_{\alpha\beta}}{2\sqrt{-g}}$ is defined in the usual way. The totally antisymmetric symbol conventionally follows $\epsilon^{0123}=1$ and the space-time element is $ds^2=-dt^2+a(t)^2d\vec{x}^2= a(\tau)^2\left(-d\tau^2+d\vec{x}^2\right)$. The combination $\frac{M_p}{\beta} \equiv f_a$ is often denoted as the axion decay constant with $M_p$ the reduced Planck mass.

We decompose the axion into the sum of its zero mode plus fluctuations, $\phi = \varphi \left( \tau \right) + \delta \phi \left( \tau ,\, \vec{x} \right)$, and we consider the temporal gauge $A_0 = 0$. The majority of studies of this system has been performed in the so called {\it homogeneous backreaction regime}, in which the axion is treated as homogeneous, $\delta \phi \left( \tau ,\, \vec{x} \right) = 0$, while the inhomogeneities of the gauge fields are  retained. With the homogeneous axion, the temporal gauge also results in $\vec{\nabla}\cdot \vec{A}=0$, and the system is described by
\begin{align}
& \varphi'' + 2 {\cal H}  \varphi' + a^2 \, \frac{d V}{d \varphi} = a^2 \frac{\beta }{M_p} \, \left\langle \vec{E} \cdot \vec{B} \right\rangle\;, \label{eq:eom-phi}\\
& {\cal H}^2 = \frac{1}{3 M_p^2} \left[ \frac{1}{2} \varphi^{'2} +  a^2 \, V + \frac{a^2}{2} \left\langle \vec{E}^2 + \vec{B}^2 \right\rangle \right] \;, 
\label{eq:eom-H}\\
& \vec{A}'' - \nabla^2 \vec{A} - \beta \frac{\varphi'}{M_p} \, \vec{\nabla} \times \vec{A} = 0  \;,  \label{eq:eom-dA}
\end{align}
In these expressions, $\langle \dots \rangle$ denotes spatial averaging, prime denotes differentiation with respect to conformal time, and ${\cal H} \equiv \frac{a'}{a}$.  The background equations make use of electromagnetic notation for brevity and clarity, 
\begin{equation}
    \vec{E}\equiv -\frac{1}{a^2}\vec{A}'\;\;\;,\;\;\;\vec{B}\equiv\frac{1}{a^2}\vec{\nabla}\times \vec{A}\;. 
\label{E-B}
\end{equation}
even though we are explicitly not associating this gauge boson with the Standard Model U(1) photon, or hypercharge. This choice would necessitate the inclusion of fermion backreaction~\cite{Domcke:2018eki}, which we do not consider here.

To describe how the coupling to the axion zero mode $\varphi$ modifies the evolution of the gauge field, we move to Fourier space
\begin{align}
{\hat A}_i ( \tau ,\, \vec{x} ) = \int \frac{d^3 k}{\left( 2 \pi \right)^{3/2}} \, {\rm e}^{ i\vec{k} \cdot \vec{x}} \, \hat{A}_i ( \tau ,\, \vec{k} ) = \sum_{\lambda = \pm} \int \frac{d^3 k}{\left( 2 \pi \right)^{3/2}} \left[ \epsilon_i^{(\lambda)} ( \vec{k} ) A_\lambda ( \tau ,\, k ) {\hat a}_\lambda ( \vec{k} ) \, {\rm e}^{ i\vec{k} \cdot \vec{x}} + {\rm h.c.} \right] \;, 
\label{eq:A-deco}
\end{align} 
with $\lambda = \pm$ corresponding to the left-handed ($+$) and right-handed ($-$) circular polarizations. The annihilation / creation operators entering in this decomposition obey $\left[ \hat{a}_\lambda ( \vec{k} ) ,\,  {\hat a}_\sigma^\dagger \left( \vec{p} \right) \right] = \delta_{\lambda \sigma} \, \delta^{(3)} ( \vec{k} - \vec{p} )$, while the polarizations operators satisfy $\vec{k} \cdot \vec{\epsilon}^{(\lambda)} \left( \vec{k} \right) = 0$, $\vec{k} \times \vec{\epsilon}^{(\lambda)} \left( \vec{k} \right) = - \lambda i k \,  \vec{\epsilon}^{(\lambda)} \left( \vec{k} \right)$, $\vec{\epsilon}^{(\lambda)} \left( - \vec{k} \right) = \vec{\epsilon}^{(\lambda)*} \left( \vec{k} \right)$, and are normalized according to $\vec{\epsilon}^{(\lambda)*} \left( \vec{k} \right) \cdot \vec{\epsilon}^{(\lambda')} \left( \vec{k} \right) = \delta_{\lambda\lambda'}$. 

As a result, the gauge field mode functions satisfy
\begin{equation}
A_\pm'' + \left( k^2 \mp k \, \frac{\beta \, \varphi'}{M_p} \right) A_\pm = 0 \,,
\label{eq:Ak}
\end{equation} 
which is often recast into
\begin{equation}
A_\pm'' + \left( k^2 \mp 2 \xi a H k \right) A_\pm = 0 
\;,\hspace{1cm} \xi \equiv \frac{\beta \, \dot{\varphi}}{2 M_p H} \;,  
\label{eq:def-xi}
\end{equation} 
where dot denotes derivative with respect to physical time. This relation shows that, for monotonic axion evolution, one gauge circular polarization is tachyonically amplified in some momentum range. Assuming $\beta \, \varphi' > 0$, $\xi$ is positive and the amplified polarization is the $\lambda = +$ one. In the special case where $\xi$ and $H$ are precisely constant, these equations admit an exact analytic solution in terms of Whittaker functions
\begin{equation}
    A_\pm(\tau,k)=\frac{{\rm e}^{\pm \pi\xi/2}}{\sqrt{2k}} W_{\mp i \xi,1/2}\left(-2 i k \tau\right)\;.
\end{equation}
It is common in the literature to use a simpler expansion of the exact solution above, for the tachyonically enhanced polarization which is valid at large $\xi$~\cite{Anber:2006xt}
\begin{equation}
    A_+(\tau,k)\simeq \frac{1}{\sqrt{2k}}\left(\frac{-k\tau}{2\xi}\right)^{1/4}{\rm e}^{\pi\xi-2\sqrt{-2\xi k \tau}}\;\;\;,\;\;\;\frac{1}{8\xi}\ll-k\tau \ll 2\xi\;,
\label{A-constant-xiH}
\end{equation}
where the subdominant imaginary component has been omitted.

The two terms bilinear in the gauge field play an important role in the physics of axion inflation and their magnitude can be approximated in the exact constant $\xi \gg 1$ and $H$ limit as
\begin{equation}
    {\cal B} \equiv \left\langle\vec{E}\cdot\vec{B}\right\rangle\simeq -2.4\cdot 10^{-4}\frac{H^4}{\xi^4}{\rm e}^{2\pi\xi}\;\;\;,\;\;\;\left\langle\frac{\vec{E}^2+\vec{B}^2}{2}\right\rangle\simeq 1.4\cdot 10^{-4}\frac{H^4}{\xi^3}{\rm e}^{2\pi\xi}\;.
\end{equation}
The first of the two expressions, playing the role of a friction term for the axion zero mode in Eq.~\eqref{eq:eom-phi},  may be used to derive a limit on the validity of the single field slow-roll attractor. Considering that $\xi\propto \sqrt{\epsilon_\varphi}\simeq\sqrt{\epsilon_H}$ (where $\epsilon_\varphi$ and $\epsilon_H$ are, respectively, the slow-roll parameters $\epsilon_\varphi \equiv \frac{M_p^2}{2} \left( \frac{1}{V} \frac{d V}{d \varphi} \right)^2$ and $\epsilon_H \equiv - \frac{\dot{H}}{H^2}$, that coincide to first order in slow-roll), it is reasonable to assume that $\xi$ will generally grow during inflation. Assuming that, at sufficiently early times, $\xi$ is small enough, the gauge field production will not affect the evolution of the axion zero-mode. In that case the inflationary trajectory would be indistinguishable from single field slow-roll $3H\dot{\varphi}\simeq V'(\varphi)$. A deviation from the standard single-field slow roll evolution then can be estimated using the ratio of gauge field friction to Hubble friction,
\begin{eqnarray}
    \left\vert \frac{\frac{\beta}{M_p}{\cal B} }{3H\dot{\phi}}\right\vert \gtrsim\sigma\;\;\;,\;\;\;\sigma\equiv 10^{-1}/1 \;\;\;,\;\;\;({\rm Mild/Strong  \; Backreaction\;  condition}) 
    \label{eq:backreaction}
\end{eqnarray}
where $\sigma$ is a parameter that distinguishes whether the gauge backreaction term is $10\%$ of Hubble friction, or equal to it. This distinction is important because in practice, as we show in the results section, even a rather small backreaction of the order of $\sigma\simeq 0.1$ is sufficient to significantly alter the inflationary attractor with respect to the ``backreactionless" single field slow-roll trajectory. In the limit of constant $\xi$ and $H$, this condition can be written more simply as
\begin{eqnarray}
    \frac{\beta H}{M_p} \lesssim 158\, \xi^{5/2}\sqrt{\sigma}\,{\rm e}^{-\pi\xi}\;.
\end{eqnarray}

Given an inflationary trajectory which violates the limit set above before the end of inflation, the system enters the ``strong backreaction regime" which is well studied in the literature in the homogeneous approximation both analytically and numerically. This homogeneous approximation typically reveals rapid oscillations in the particle production parameter $\xi$ in the strong backreaction regime, which can be traced back to the time-delayed gauge field friction arising from Eqs.~\eqref{eq:eom-phi} and \eqref{eq:eom-dA}~\cite{DallAgata:2019yrr,Domcke:2020zez,Peloso:2022ovc,Gorbar:2021rlt}. Such oscillations could lead to distinct
signatures in the spectra of gravitational waves and density perturbations~\cite{Garcia-Bellido:2023ser,vonEckardstein:2025oic,Barbon:2025wjl}. However, as we will analyze in the next section, these features have been called into question when nonlinear physics, arising by the inhomogeneities of the axion, are taken into account~\cite{Figueroa:2023oxc}.
%
\section{The role of axion inhomogeneities}
\label{sec:axion-inflation-nonlinear}
%
The homogeneous backreaction approximation presented in the previous section has proven highly successful in describing the dynamics of axion inflation in the regime of weak backreaction. However, recent lattice simulations have revealed substantial departures from this picture due to dynamical formation of sizable axion gradients which, depending on the axion potential, can invalidate the approximation of the previous section~\cite{Caravano:2021bfn,Caravano:2022epk,Figueroa:2023oxc,Caravano:2024xsb,Figueroa:2024rkr,Lizarraga:2025aiw,Jamieson:2025ngu}. The lattice simulations, while providing the full non-perturbative answer, require considerable computational resources and are hence currently unsuitable to systematically investigate the parameter space of axion inflation. It is thus crucial to develop faster, approximative methods with a well-defined regime of validity, which can guide the lattice resources to the phenomenologically most relevant parts of the parameter space.

To understand the development of these large gradients, one can consider the linearized equation for the axion fluctuations arising from their interaction with the gauge field 
\begin{align}
& \delta\phi''+2{\cal H}\delta\phi' -\vec{\nabla}^2 \delta\phi + a^2 \, \frac{d^2 V}{d\varphi^2} \, \delta\phi = - \frac{a^2\beta}{4 M_p} F_{\mu \nu} {\tilde F}^{\mu \nu} \;. 
\label{eq:eom-dphi}
\end{align}

We see that the coupling results in a transfer of energy between axion fluctuations and the gauge field, which, depending on the inflaton potential and the value of the coupling $\beta$, can become significant.\footnote{We instead can ignore scalar metric perturbations which provide gravitational interactions, which are subdominant to those considered here since they are not sourced directly by the tachyonically enhanced gauge fields~\cite{Barnaby:2011vw}.} The resulting enhancement of axion fluctuations then mediates the redistribution of gauge-field power across momentum space (in a U(1) theory, interactions between gauge modes are mediated by axion gradients). Accounting for these effects will result in an improved evaluation of the backreaction term, so that Eq.~\eqref{eq:eom-phi} can be replaced by
\begin{equation}
    \varphi'' + 2 {\cal H}\,\varphi' + a^2 \frac{dV}{d\varphi}
    = a^2 \frac{\beta}{M_p}  \left( {\cal B}^{(0)} + \delta^{(1)} {\cal B} \right)  \,,
    \label{eq:improved_backreaction}
\end{equation}
where $\delta^{(1)} {\cal B}$ schematically denotes a perturbative correction to the homogeneous backreaction term ${\cal B}^{(0)}$, as discussed below.

The impact of the modified backreaction on the axion evolution is typically more significant than that on the Friedmann equation, since the driving force in the equation of motion of the inflaton is slow-roll suppressed, and it is therefore more sensitive to backreaction. For this reason, we expect that homogeneous backreaction will adequately describe the background evolution only as long as
\begin{equation}
\left\vert \delta^{(1)} {\cal B} \right\vert \; \ll \;\left\vert {\cal B}^{(0)} \right\vert \;. 
\label{criterion}
\end{equation}
This is our proposed criterion for the validity of homogeneous backreaction. 

The central question may then be reformulated as follows: \textit{How can we perform a perturbative calculation with a well-defined regime of validity, such that within this regime, we can self-consistently determine the corrected backreaction in the presence of axion gradients?} To address this question, we propose two distinct methods for evaluating $\delta^{(1)} {\cal B}$, and then, the criterion \eqref{criterion}. The first method is based on the GEF approximation, which has proven to be both an effective and computationally efficient framework for studying strong backreaction within the homogeneous backreaction approximation, and which has also been extended to incorporate the effects of axion fluctuations perturbatively. The second method is based on a loop expansion of the backreaction, in which the leading contribution is evaluated at one loop using the in-in formalism.
%
\subsection{Method 1: Perturbatively including inhomogeneities in the GEF}
\label{sec:criterion1}
%
The relevance of inhomogeneities in the axion field on the gauge field backreaction, i.e.\ on the right-hand side of Eq.~\eqref{eq:improved_backreaction}, can be self-consistently evaluated by perturbatively including the axion gradient terms using the formalism developed in Ref.~\cite{Domcke:2023tnn}. This formalism is an extension of the gradient expansion formalism (GEF) introduced for axion inflation in Refs.~\cite{Gorbar:2021rlt,Gorbar:2021zlr,vonEckardstein:2023gwk}. We briefly review the key concepts here, with the goal of introducing a first-order correction to $\langle \vec E \vec B \rangle$, which serves as a measure for the impact on the backreaction.

The GEF trades the equations for the individual gauge field modes~\eqref{eq:Ak} for a set of coupled first order ordinary differential equations evolving the derivatives of the two-point functions,
\begin{align}
	\mathcal{P}_{X}^{(n)} &= \frac{1}{a^n}\left\langle \vec{X}\cdot(\vec{\nabla}\times)^n \vec{X}\right\rangle\,,
	\quad
	\mathcal{P}_{XY}^{(n)} = -\frac{1}{a^n}\left\langle \vec{X}\cdot(\vec{\nabla}\times)^n \vec{Y}\right\rangle\,,
	\label{eq:2pt}
\end{align}
with $\{X,Y\} = \{E,B\}$.
Complemented with the evolution of the inflaton field,  Eqs.~\eqref{eq:eom-phi} and \eqref{eq:eom-dphi}, and the Hubble parameter, Eq.~\eqref{eq:eom-H}, this provides an equivalent description of the dynamics in the limit that axion gradient terms can be neglected. The tower of equations for ${\cal P}^{(n)}$ can be truncated at finite order of the derivatives while yielding an accurate result for the evolution of the the inflaton field and gauge field power spectra. See Refs.~\cite{Gorbar:2021rlt,vonEckardstein:2023gwk} for details.

The GEF formalism can be extended to perturbatively include axion gradients as long as these remain sufficiently small to ensure perturbative control. Concretely, this amounts to extending the lowest order equations of the GEF tower ($n = \{0,1\}$) to include 3-point functions including one power of either $\dot{\delta \phi}$ or $\vec \nabla \delta \phi$~\cite{Domcke:2023tnn}. To leading order, this takes into account the modification of the gauge field spectra, and in particular the gauge field backreaction $\langle E B \rangle$ due to presence of axion fluctuations. It is this change in the gauge field backreaction that in turn leads to significant changes in the evolution of the coupled axion gauge field system.

This approach is by construction perturbative and breaks down once the axion gradients become too large. As demonstrated in Ref.~\cite{Domcke:2023tnn} by comparison with lattice simulations, this occurs when the axion gradient energy reaches about $5\%$ of the axion kinetic energy. While a study of the phenomenology beyond this point is out of reach of this perturbative method, it nevertheless provides a self consistent analysis tool to determine the regime of validity of the perturbative evolution.

In this paper we use GEF expansion supplemented with a perturbative expansion in the axion gradients to determine the correction that these gradients provide to the backreaction term $\langle \vec E \cdot \vec B \rangle$~\cite{Domcke:2023tnn}. Concretely, we solve the GEF tower up to $n_{\mathrm{max}} = 150$,\footnote{We checked that our results are insensitive to the exact choice of $n_{\mathrm{max}}$ for $n_{\mathrm{max}} \geq 150$.} including axion gradients to first order as described in Ref.~\cite{Domcke:2023tnn}. More specifically, the explicit set of equations we solve is the set presented in Appendix A.1 of Ref.~\cite{Domcke:2023tnn}.
This yields an improved backreaction term ${\cal B}^{(0)} + \delta^{(1)}_{\delta \phi \text{GEF}} {\cal B}$ in the sense of Eq.~\eqref{eq:improved_backreaction}, with the subscript ${\delta \phi}$GEF indicating the perturbative prescription used to include the impact of the axion gradients in the gauge field backreaction.
%
\subsection{Method 2: Nonlinear correction to the gauge field backreaction with the in-in formalism.}
\label{sec:criterion2}
%
Within homogeneous backreaction, combining Eqs. \eqref{E-B} and \eqref{eq:A-deco} the backreaction term is given by 
\begin{eqnarray} 
{\cal B}^{(0)} \left( \tau \right) & \equiv & \left\langle \left( \vec{E} \cdot \vec{B} \right)^{(0)} \left( \tau \,, \vec{x} \right) \right\rangle = - \frac{1}{2 a^4 \left( \tau \right)} \int \frac{d^3 k}{\left( 2 \pi \right)^3} k \, \frac{d}{d \tau} \left\vert A \left( \tau ,\, k \right) \right\vert^2 \;, 
\label{eq:B0}
\end{eqnarray} 
where $A \left( \tau ,\, k \right)$ denotes the gauge mode function amplified according to Eq.~\eqref{eq:Ak}. We evaluate the departure of the full system from \eqref{eq:B0} via the in-in formalism
\begin{equation}
\delta^{(1)}_\text{in-in} {\cal B} \left( \tau \right) = - \int^\tau d \tau_1 \int^{\tau_1} d \tau_2 \left\langle \left[ \left[ \vec{E} \left( \tau ,\, \vec{x} \right) \cdot \vec{B} \left( \tau ,\, \vec{x} \right) ,\, H_{\rm int} \left( \tau_1 \right) \right] ,\, H_{\rm int} \left( \tau_2 \right) \right] \right\rangle  \;.  
\label{eq:B1-inin}
\end{equation} 
To obtain the interaction Hamiltonian, we start from the action describing the system within homogeneous backreaction: 
\begin{equation}
    S^{(0)}=\int d^4 x \sqrt{-g}\left[\frac{M_p^2}{2}R-\frac{1}{2}\partial_\mu\phi\partial^\mu \phi-V(\phi)-\frac{1}{4}F_{\mu\nu}F^{\mu\nu}-\beta\frac{\varphi}{4 M_p}F_{\mu\nu}\tilde{F}^{\mu\nu}\right] \;, 
\label{action-0}
\end{equation}
which differs from the full action \eqref{action} by the fact that only the axion zero mode is coupled to the gauge field. The difference between the two actions $S - S^{(0)} = \int d^4 x \, \sqrt{-g} \; {\cal L}_{\rm int}$ provides the interaction hamiltonian 
\begin{equation}
H_{\rm int} = - \int d^3 x \sqrt{-g} \; {\cal L}_{\rm int} = \frac{\beta}{8 M_p} \int d^3 x \;  \delta \phi \, \epsilon^{\mu \nu \alpha \beta} \, F_{\mu \nu} \, F_{\alpha \beta} \;.
\label{Hint}
\end{equation}

We expect that the axion evolution is well described by homogeneous backreaction as long as the correction \eqref{eq:B1-inin} is much smaller than the `unperturbed' term \eqref{eq:B0}. This mirrors the criteria for perturbativity evaluated in Ref.~\cite{Peloso:2016gqs}, where analogous computations were performed for the $\left\langle A^2 \right\rangle$ and $\left\langle \delta \phi^2 \right\rangle$ correlators. Contrary to that work, here we study the correlator that is directly related to backreaction. Moreover, we evaluate Eqs. \eqref{eq:B0} and \eqref{eq:B1-inin} using `unperturbed' mode functions obtained from the numerical evolution of the system of eqs.~\eqref{eq:eom-phi},~\eqref{eq:eom-H}, and~\eqref{eq:eom-dA}. This improves over the computations of Ref.~\cite{Peloso:2016gqs} that employed analytical mode function \eqref{A-constant-xiH}, valid only for constant or adiabatically evolving $H$ and $\xi$. We stress that 'unperturbed' in this language refers to the assumption of homogeneous backreaction.

Details of the computation of the expression~\eqref{eq:B1-inin} are given in App.~\ref{app:loop}. Omitting for brevity the momentum dependence, as well as time and spatial derivatives (needed to form the `electric' and `magnetic' contribution that we are evaluating, as well as those contained in the interaction), the terms have the structure 
\begin{align}
\begin{split}
& \left\langle \left[ \left[ A \left( \tau \right) A \left( \tau \right) ,\, \delta \phi \left( \tau_1 \right) A \left( \tau_1 \right) A \left( \tau_1 \right) \right] ,\, \delta \phi \left( \tau_2 \right) A \left( \tau_2 \right) A \left( \tau_2 \right) \right] \right\rangle \\
& \quad \simeq \left[ A \left( \tau \right) ,\, A \left( \tau_1 \right) \right] \left[ \delta \phi \left( \tau_1 \right) ,\, \delta \phi \left( \tau_2 \right) \right] \left\langle A \left( \tau_1 \right) A \left( \tau_2 \right) A \left( \tau_2 \right) A \left( \tau \right) \right\rangle \;, 
\label{correlator}
\end{split}
\end{align} 
where in the second expression we have retained only the dominant contribution, proportional to the expectation value of four gauge fields. 

\begin{figure}
\centering
\includegraphics[width=0.99\linewidth]{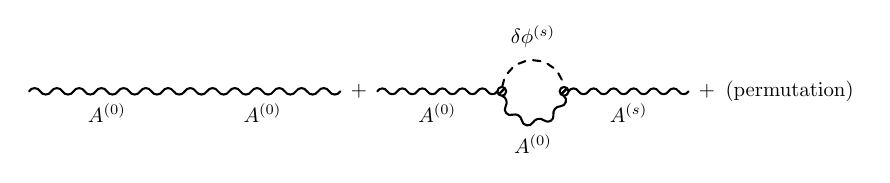}
\caption{Diagrammatic interpretation of the leading order correction to $\langle \vec E \vec B \rangle$ in the in-in formalism due to axion fluctuations. The superscript `$(0)$' refers to solutions under homogeneous backreaction, while `$(s)$' indicates fields sourced through the axion gauge field coupling. The two diagrams correspond to the two terms \eqref{eq:B0} and \eqref{eq:B1-inin}. Temporal and spatial derivatives (as needed to obtain the `electric' and `magnetic' contributions) are omitted.} 
\label{fig:0+1loop}
\end{figure}

This term contributes to the one loop diagram visualized in Fig.~\ref{fig:0+1loop} and it is interpreted as the interference term between one `unperturbed' (amplified) $A^{(0)}$ gauge mode (the last term in \eqref{correlator}) and a gauge mode modified at one loop by the interaction of one `unperturbed' $A^{(0)}$ gauge mode and one `sourced' scalar field mode. 

In App.~\ref{app:loop} we evaluate the correlator 
\begin{align}
\left\langle \left[ \left[ {\hat A} \left( \tau ,\, \vec{k}_1 \right) \; {\hat A} \left( \tau' ,\, \vec{k}_2 \right) ,\, H_{\rm int} \left( \tau_1 \right) \right] ,\, H_{\rm int} \left( \tau_2 \right) \right] \right\rangle \equiv {\cal C}^{(0,0)} \left( \tau ,\, \tau' ,\, \tau_1 ,\, \tau_2 ,\, k_1 \right) \delta^{(3)} \left( \vec{k}_1 + \vec{k}_2 \right) \,. 
\label{correlator-C0}
\end{align}
The result, reported in Eq.~\eqref{Atau-Ataup} is symmetric in the two external times $\tau$ and $\tau'$. 
Defining 
\begin{equation}
{\cal C}^{(m,n)} \left( \tau ,\, \tau' ,\, \tau_1 ,\, \tau_2 ,\, k_1 \right) \equiv \frac{\partial^m}{\partial \tau^m}  \frac{\partial^n}{\partial \tau^{n'}} \; {\cal C}^{(0,0)} \left( \tau ,\, \tau' ,\, \tau_1 ,\, \tau_2 ,\, k_1 \right) 
\end{equation} 
the expression \eqref{eq:B1-inin} evaluates to 
\begin{align} 
\begin{split}
\delta^{(1)}_\text{in-in} {\cal B} \left( \tau \right)  &=  \frac{1}{a^4} \int \frac{d^3 k_1 d^3 k_2}{\left( 2 \pi \right)^3} \; {\rm e}^{i \left( \vec{k}_1 + \vec{k}_2 \right) \cdot \vec{x}} \epsilon_i \left( {\hat k}_1 \right) \epsilon_i \left( {\hat k}_2 \right) \delta^{(3)} \left( \vec{k}_1 + \vec{k}_2 \right) \\
& \int^\tau d \tau_1 \int^{\tau_1} d \tau_2 \; \left[ \frac{k_2}{2} \; {\cal C}^{(1,0)} \left( \tau ,\, \tau ,\, \tau_1 ,\, \tau_2 ,\, k_1 \right) + \frac{k_1}{2} \; {\cal C}^{(0,1)} \left( \tau ,\, \tau ,\, \tau_1 ,\, \tau_2 ,\, k_1 \right) \right] \;. 
\end{split}
\end{align} 
After performing some momentum integrations, and exploiting the symmetry over the external times, this expression simplifies to
\begin{eqnarray} 
\delta^{(1)}_\text{in-in} {\cal B} \left( \tau \right) &=& \frac{1}{a^4} \int d k \; \frac{k^3}{2 \pi^2} \; \int^\tau d \tau_1 \int^{\tau_1} d \tau_2 \; {\cal C}^{(1,0)} \left( \tau ,\, \tau ,\, \tau_1 ,\, \tau_2 ,\, k \right) \;. 
\label{dB1-final} 
\end{eqnarray} 
In the next section we evaluate this quantity with the full time dependence of the mode functions and background fields taken into account. 

\subsection{Comparison of the two methods}

Both methods target the same physical quantity: the correction to the backreaction term ${\cal B}$ due to sourced axion modes, i.e.\ axion gradients, while they differ in the perturbative expansion scheme used (see above). A further difference concerns the implementation of $\delta^{(1)}{\cal B}$ within the respective framework. In the $\delta\phi$GEF method, $\delta^{(1)}{\cal B}$ is consistently included when evaluating the dynamical evolution of the system, whereas in the in-in formalism,  $\delta^{(1)}{\cal B}$ serves as a diagnostic tool but does not enter the evaluation of the dynamics of the axion gauge field system at this order. As we will see below, while both methods reliably detect any meaningful change in the backreaction ${\cal B}$, the $\delta\phi$GEF method also accounts for its overall impact on the evolution. As we will see, in most cases this subtle difference is irrelevant. However, for example in cases of mild backreaction, the impact on ${\cal B}$ may be relevant, but nevertheless subdominant compared to Hubble friction. In this sense, we consider the evaluation in the in-in formalism a sufficient, though possibly not necessary criterion for the validity of homogeneous backreaction.

We stress that this is not an intrinsic limitation of the in-in computation. It could be refined, for example, by extending the in-in computation to two loops (which, for technical reasons we describe below, would provide a much better indicator of the IR to UV transfer) or by measuring also the modification of the shape of the spectral backreaction and not only the overall integral.

\section{Results}
\label{sec:results}

Having now at our disposal the tools developed in the previous section, we proceed to evaluate the criterion~\eqref{criterion} for different realizations of axion inflation in the strong backreaction regime. In order to make our analysis concrete and connect with past literature, we  restrict it to the same initial conditions and inflationary potential as in past related works~\cite{Figueroa:2023oxc,Domcke:2023tnn,Figueroa:2024rkr}. Namely, we opt for a quadratic potential, with the mass and initial field value such that the power of scalar fluctuations is compatible with observations for scales leaving the horizon $N_{\rm tot}\simeq 60$ e-foldings before the end of inflation.\footnote{This is the typical number of e-foldings between the CMB production and the end of inflation required for a quadratic potential. In our analysis, we treat this only as a convenient reference value since (i) the total amount of inflation starting from a given value $\varphi_{\rm ini}$ depends on the subsequent amount of backreaction, and (ii) matching the CMB predictions is not a goal of the present work (as evident also by the choice of the potential). Similarly, we do not aim for a consistent description of the reheating phase, which for the potential and couplings considered here would likely overproduce GWs~\cite{Adshead:2019lbr,Adshead:2019igv}.}
\begin{eqnarray}
 V(\varphi)=\frac{1}{2}m^2\varphi^2   \;\;\;,\;\;\; m = 6.16\cdot 10^{-6}\; M_p\;\;\;{\rm and}\;\;\; \varphi_{\rm ini}=-15.55 \; M_p\;,
\end{eqnarray}
where the subscript ``ini" indicates that the corresponding variable or field is evaluated at the initial moment. The initial velocity of the axion is imposed by the single field slow-roll attractor $3H\dot{\varphi}_{\rm ini}\simeq - dV(\varphi)/d\varphi\Big|_{\varphi=\varphi_{\rm ini}}$, since the couplings $\beta$ under consideration will be small enough for this to be an excellent approximation at CMB scales. We conventionally define $a_{\rm ini}\equiv 1$ and use the number of e-foldings as our time variable defined as $N\equiv \ln\left(\frac{a}{a_{\rm ini}}\right)$.

The results of our two different methods to evaluate the correction to the backreaction are displayed in two sets of complementary plots. Here we give a general description of these two sets, describing the various lines displayed in them. The results shown in each individual figure are then discussed in the next two subsections.

\bigskip 
{\bf First set of Figures}. The first set of figures (specifically, Figs.~\ref{fig:xi-lattice} and~\ref{fig:xi-large-coupling}) displays the evolution of the production parameter, $\xi$ defined in Eq.~\eqref{eq:def-xi}, evaluated with different levels of approximation, specified in the caption of Fig.~\ref{fig:xi-lattice}. The results of method 1 are primarily displayed in these plots via a red line that tracks the real time evolution of $\xi$ taking into account axion inhomogeneities to first order via the GEF formalism outlined in Sec.~\ref{sec:criterion1}. Since our goal is to identify the moment in time for which axion gradients modify the backreaction term appreciably, we mark with a star symbol \ding{72} the moment at which the correction to the backreaction becomes greater than $10\%$ of the term evaluated within homogeneous backreaction, namely when the time at which
\begin{align}
\delta^{(1)} {\cal B} = 0.1 \; {\cal B}^{(0)} \;. 
\label{dB-B-01}
\end{align}
The red (respectively, black) mark refers to the correction $\delta^{(1)}_{\delta \phi \text{GEF}} {\cal B}$ evaluated as outlined in Sec.~\ref{sec:criterion1} (respectively, $\delta^{(1)}_\text{in-in} {\cal B}$, evaluated as outlined in Sec.~\ref{sec:criterion2}). Finally, the vertical solid line (separating the white and gray portion of each plot) indicates the moment at which the axion gradient energy reaches $1\%$ of the axion kinetic energy, while the vertical dashed line shows the moment at which the ratio between the gradient and kinetic energies is $5\%$, see Ref.~\cite{Domcke:2023tnn}.

\bigskip 
{\bf Second set of Figures}. 
The second set of figures (specifically, Figs.~\ref{fig:spec-15} and~\ref{fig:spec-50}, as well as Figs.~\ref{fig:spec-18}--\ref{fig:spec-25} in App.~\ref{app:spectra}) serves to showcase the results of method 2,  by showing the normalized \textit{spectral backreaction} computed via the in-in formalism developed in Sec.~\ref{sec:criterion2}. We define the normalized spectral backreaction contributions $\tilde{{\cal B}}^{(0)} \left(N, k \right)$ and $\delta^{(1)} \tilde{{\cal B}} \left(N, k \right)$ as 
\begin{eqnarray}
\frac{\beta\,{\cal B}^{(0)} \left( N \right)}{M_p V'(\varphi)}\equiv \int d\ln k \; \tilde{{\cal B}}^{(0)} \left(N, k \right) \;\;\;,\;\;\; \frac{\beta\,\delta^{(1)} {\cal B} \left( N \right)}{M_p V'(\varphi)}\equiv \int d\ln k \; \delta^{(1)} \tilde{{\cal B}} \left(N, k \right) \;, 
\label{def-Btilde}
\end{eqnarray}
where ${\cal B}^{(0)} \left( N \right)$ and  $\delta^{(1)} {\cal B} \left( N \right)$ are, respectively, the homogeneous backreaction term~\eqref{eq:B0} and the 
first order correction~\eqref{eq:B1-inin}. These spectra are plotted for each run, with $\beta=\{15,18,20,25,50\}$, at different times, where the time, expressed in e-foldings $N$ and the corresponding value of $\xi$ are labeled above the panels. As can be verified a posteriori, the spectra for the first four cases are qualitatively similar to each other, and therefore in order to avoid cluttering the main text, we display the spectra for the three intermediate cases in App.~\ref{app:spectra}, keeping only the spectra of the lowest and greatest coupling in the main text. The times of the various panels are chosen so that the earliest (respectively, latest) panels feature a small (respectively, significant) correction term. Additionally, we display the scale that crosses the horizon with a vertical green line at each panel and also display the greatest wave number that has ever been tachyonic in the entire history of the evolution up to the moment plotted in red 
\begin{eqnarray}
    k_H(N) \equiv&  a(N) H(N)\;\;\;\;\;\;\;\;\;\;\;\;\;\;\;\;\;\;\;\;\;\;\;\;\;\;\;\;\;\;\;\;\;\;\;\;&({\rm green\; line})\;,\\
    k_{\rm max}(N)\equiv&  {\rm MAX}\left[2\xi(N') a(N') H(N')\right]\Big|_{N'\leq N} \;\;\;&({\rm red\; line})\;.
\end{eqnarray}
A horizontal blue line at unity serves as a visual reference: when the spectral backreaction, shown by the black and brown curves, approaches this line, it signals the onset of strong backreaction.
%

\subsection{Lattice validation of the criteria and physical interpretation}
\label{sec:lattice-validation}
\begin{figure}
    \centering
    \includegraphics[width=0.32\linewidth]{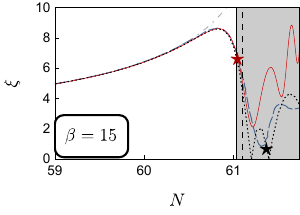} \hfill
    \includegraphics[width=0.32\linewidth]{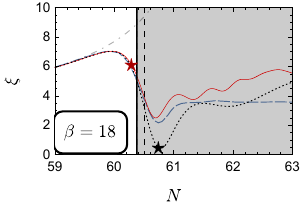} \hfill
    \includegraphics[width=0.32\linewidth]{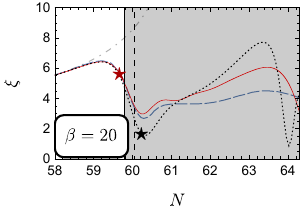}
    \caption{Evolution of the particle production parameter $\xi$ as a function of the number of e-folds $N$ at varying degrees of approximation for different values of the axion gauge field coupling $\beta$. The grey dot-dashed line displays the evolution in the absence of backreaction; the black dotted line considers the homogeneous approximation, the red solid line is the GEF result where axion inhomogeneities are included perturbatively. Finally, the lattice result~\cite{Figueroa:2023oxc}, capturing the full non-perturbative evolution, is shown in dashed blue. The two stars indicate the moments at which the two evaluations of the criterion~\eqref{criterion} are triggered, indicating when the axion inhomogeneities start affecting the backreaction term significantly. The vertical solid (respectively, dashed) line indicates the time at which the axion gradient energy becomes $1\%$ (respectively, $5\%$) of the axion kinetic energy.}
    \label{fig:xi-lattice}
\end{figure}
\begin{figure}
    \centering
    \includegraphics[width=0.95\linewidth]{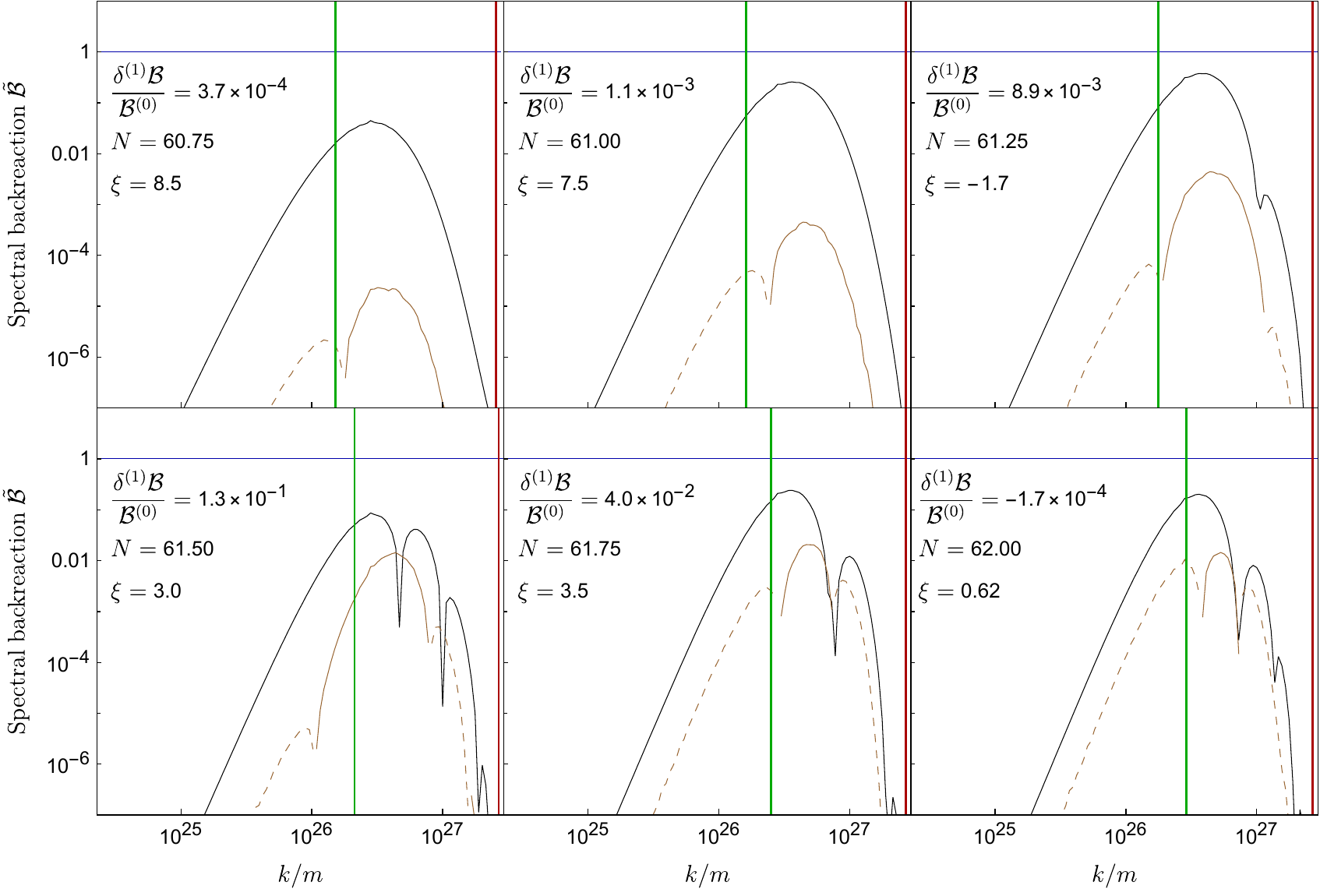}
    \caption{Normalized spectral backreaction for coupling $\beta=15$. 
    Different panels refer to different times, with the corresponding value of $\xi$ obtained within homogeneous backreaction. The black (respectively, brown) line shows the integrand in Eq.~\eqref{def-Btilde} evaluated within the homogeneous backreaction approximation (respectively, the first order correction evaluated with the in-in formalism). The correction is shown with a solid line when positive, and with a dashed line when negative. The green and red vertical lines denote, respectively, the comoving horizon at the moment $N$ which the panel refers to and the greatest wave number that has ever been tachyonic in entire history $N' \leq N$.
   }
    \label{fig:spec-15}
\end{figure}

In this subsection we validate our criteria for the validity / failure of homogeneous backreaction by comparing them against lattice simulations present in the literature~\cite{Figueroa:2023oxc,Figueroa:2024rkr}, and by the more immediate criterion first advocated in~\cite{Domcke:2023tnn} of comparing the gradient vs.\ the kinetic energy of the axion. The comparison is performed via a series of plots whose content has been described in the first part of this section.

The three panels of Fig.~\ref{fig:xi-lattice} refer to three different (and, progressively, increasing) values of the axion-gauge coupling. As described above, in each panel we superimpose the evolution of the particle-production parameter obtained from the fully inhomogeneous, nonperturbative  lattice calculation of~\cite{Figueroa:2023oxc}, together with the result of the linear approximation, which neglects axion inhomogeneities, and the GEF computation, which incorporates them perturbatively. This provides an important calibration for our two criteria implementations of the criterion, and guides our choice of which threshold values to use in both cases. 

Focusing first on $\delta^{(1)}_{\delta \phi \text{GEF}} {\cal B}$, we see that the red star in Fig.~\ref{fig:xi-lattice} consistently lie close to the time at which the homogeneous backreaction result (black dotted line) begins to deviate from the lattice result (blue dashed line). This provides nontrivial support for the criterion in the benchmark cases considered. We also note that this time systematically lies near the vertical lines indicating placed at the $1\%$ and $5\%$ ratios between the gradient and kinetic axion energies, indicating that the criterion proposed here is consistent with that suggested in Ref.~\cite{Domcke:2023tnn}.

Turning now to $\delta^{(1)}_\text{in-in} {\cal B}$, we again find that the black stars lie close to the time at which the homogeneous approximation departs from the lattice result, providing a complementary and consistent handle on the importance of axion inhomogeneities. Fig.~\ref{fig:spec-15} (see also Figs.~\ref{fig:spec-18} and \ref{fig:spec-20} in App.~\ref{app:spectra}) offer further insight into how axion perturbations modify the strong-backreaction regime. A first important observation is that the one-loop correction to the backreaction exhibits a momentum-space structure that peaks at roughly the same scales as the homogeneous backreaction result. This is consistent with the fact that, at one loop, the dominant contribution is an interference term between a tachyonically enhanced vacuum gauge mode and a sourced gauge mode whose amplitude has been modified by the presence of axion gradients. Because one of the two external legs is evaluated as a vacuum mode, the one-loop correction has support only up to the maximum wave number \(k_{\rm max}(N)\): by construction, the modes are not enhanced beyond that scale, and they are hence set to zero in our prescription.

A second notable feature is that, especially at early times when backreaction is still mild, the correction is negative at small wave numbers and positive at larger ones. Since the full backreaction is given by the sum of the homogeneous contribution and the correction, this sign pattern suggests a redistribution of power in the gauge field from lower to higher momenta. We interpret this as evidence for an incipient transfer of power from the IR toward the UV. This behavior is qualitatively reminiscent of nonlinear bosonic systems with large occupation numbers in the infrared, such as those encountered in studies of preheating~\cite{Micha:2004bv,Podolsky:2005bw,Adshead:2015pva,Cuissa:2018oiw}. At the same time, we emphasize that our calculation does not establish a fully developed cascade in the technical sense, since the perturbative nature of our computation cannot track a sustained spectral flux or self-similar transport over a wide dynamical range.

The physical intuition for the direction of this transfer is straightforward. In the examples we study, the Chern--Simons coupling is large enough that significant backreaction develops before the end of inflation. As a result, a substantial amount of energy is deposited into gauge bosons with large occupation numbers, concentrated in a relatively narrow band of wave numbers near horizon crossing. Once our criteria are triggered, nonlinear effects become important and can redistribute this power across momentum space. In that situation, it is natural to expect the spectrum to broaden toward higher physical momenta.  This occurs because the  spectrum of the modes $A^{(0)}$ (those generated as long as the axion  gradients are negligible) is much more populated in the infrared (IR) with respect to a thermal spectrum with the same energy, so interactions will be characterized by processes with fewer quanta out than in, that move the spectrum toward the ultraviolet (UV). The perturbative process corresponding to the second diagram in Fig.~\ref{fig:0+1loop} is one such example, since the right external leg is the result of an $A^{(0)}$ mode interacting with an intermediate sourced axion mode, which has been sourced by two modes $A^{(0)}$ (therefore, three $A^{(0)}$ modes concur to the formation of $A^{(s)}$). We note, however, that by construction, the IR-to-UV transfer captured by this interference term is necessarily limited, since it does not have support beyond that of the $A^{(0)}$ modes. We expect the effect to be more pronounced at two loops, where correlators involving two sourced gauge modes can contribute. In that case, momentum-space convolutions of sourced modes can generate support beyond the one-loop cutoff, making the transfer of power to larger wave numbers more manifest. The existing lattice simulations show this clear shift of the gauge modes spectrum when axion gradients are included: see for instance Fig.~2 of~\cite{Figueroa:2023oxc}.

We therefore conclude that the two criteria are consistent with one another in pinpointing when axion inhomogeneities become important, and that both are supported by the available lattice simulations. In the next subsection, we move beyond the portion of parameter space for which lattice results currently exist.
\subsection{Exploring stronger couplings}
\label{sec:strong-coupling}
\begin{figure}
    \centering
    \includegraphics[width=0.45\linewidth]{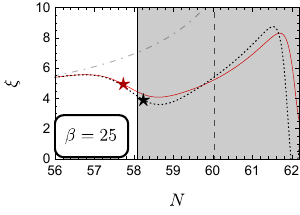}
    \includegraphics[width=0.45\linewidth]{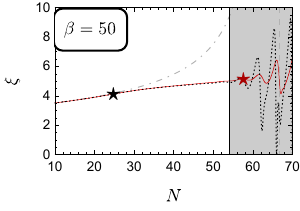}
    \caption{The evolution of the particle production parameter $\xi$ for larger couplings. The curves and markers are defined as in Fig.~\ref{fig:xi-lattice}.}
    \label{fig:xi-large-coupling}
\end{figure}
\begin{figure}
    \centering
\includegraphics[width=0.95\linewidth]{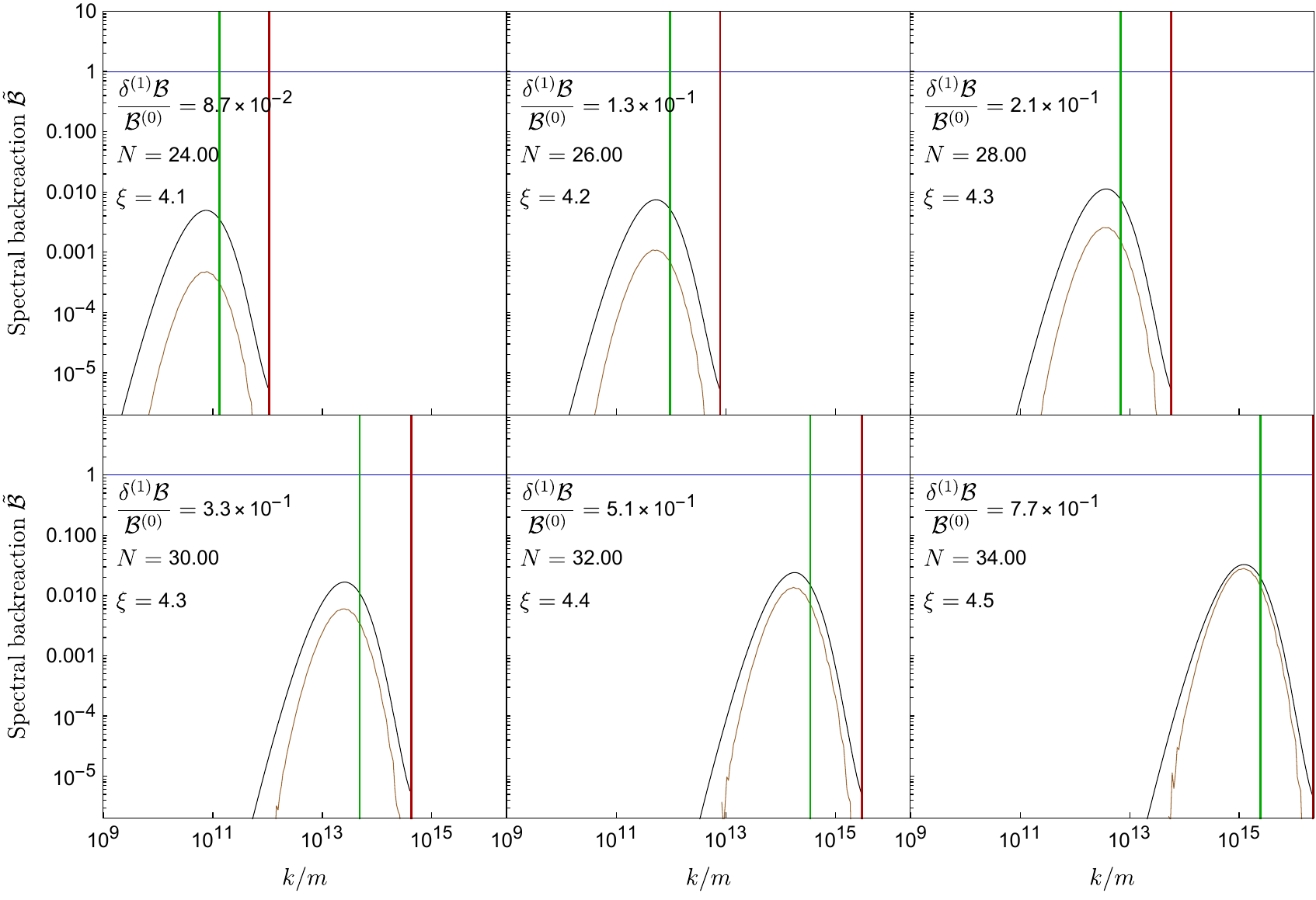}
    \caption{Normalized spectral backreaction for coupling $\beta=50$. The black line displays the homogeneous backreaction approximation~\eqref{eq:B0} while the brown line shows the correction to the backreaction~\eqref{eq:B1-inin} because of the presence of inhomogeneities. Solid lines correspond to a positive spectrum whereas dashed lines to negative values. Contrary to all the cases above, there is no sign change in central region of the brown curve, showing no transfer of power from IR to UV in the gauge modes. This relates to the qualitatively different evolution with respect to the previous cases shown in the right panel of Fig.~\ref{fig:xi-large-coupling}, as we discuss in the text. 
    }
    \label{fig:spec-50}
\end{figure}

The analysis presented in the previous section shows that both criteria developed in this work yield plausible results that are broadly consistent with existing nonperturbative lattice simulations. This validation is essential for establishing the reliability of our criteria before extrapolating to larger couplings, where lattice results are not yet available. In this section, we extend the analysis to larger values of the coupling and consider, as representative examples, the cases \(\beta = 25\) and \(\beta = 50\). We follow the same methodology as in the previous examples, and present the corresponding results in Fig.~\ref{fig:xi-large-coupling}.

The first example, corresponding to $\beta = 25$, coincides with the largest coupling considered in the previous study of Ref.~\cite{Domcke:2023tnn}. As in the earlier examples, we find that the two criteria remain in good agreement with each other, indicating that the onset of axion-gradient effects occurs at comparable times and, in particular, close to the point at which the axion gradient energy reaches approximately $1\%$ of the axion kinetic energy. One notable feature of this case is that the evolution of the particle production parameter $\xi$ remains closely correlated between the GEF approximation including inhomogeneities (red line) and the homogeneous backreaction approximation (black line), with the maxima and minima occurring at nearly the same times. This conclusion is also broadly supported by the spectral backreaction obtained from the in-in computation (see Fig.~\ref{fig:spec-25} in App.~\ref{app:spectra}). Both methods suggest a brief interval of strong backreaction which seems to be under perturbative control, with only a modest impact of the axion gradients. This regime extends approximately from $N \simeq 58$ to $N \simeq 59.5$.

Does this point to a more general phenomenon at large couplings? We address this question in the right panel of Fig.~\ref{fig:xi-large-coupling} and in Fig.~\ref{fig:spec-50}, where results are presented for $\beta = 50$.\footnote{Such a large coupling, combined with the quadratic potential, results in $\xi_{\rm CMB}\simeq 3.2$ which is incompatible with bounds on scalar nongaussianity~\cite{Barnaby:2011vw}. However, we are operating under the assumption of a toy model in this work, in order to explore the inflationary dynamics in these higher couplings.} For this value, the evolution of $\xi$  exhibits several striking features that were largely absent in the previous cases. Firstly, we observe a significant departure of the trajectories including backreaction effects (red and black lines) from the single-field slow-roll solution shown by the grey dot-dashed line. This corresponds to a regime in which the friction induced by backreaction remains relevant over a sufficiently prolonged interval to generate a substantial deviation from the single-field slow-roll trajectory, while still remaining well below Hubble friction. This is precisely the regime defined in Eq.~\eqref{eq:backreaction} and referred to as \textit{mild backreaction}.

More specifically, the threshold for mild backreaction is crossed at approximately \(N \simeq 27.6\), while the threshold for strong backreaction is crossed at \(N \simeq 47.2\), both defined according to Eq.~\eqref{eq:backreaction} evaluated under the homogeneous backreaction approximation. This extended period of mild backreaction is characterized by a relatively large value of the particle production parameter, which could in principle be associated with observable gravitational waves at interferometer scales. Moreover, \(\xi\) does not exhibit oscillatory features during this phase. To the best of our knowledge, this is the first example of a dynamical realization of such a regime in the literature.\footnote{Note that the recent stability analysis of~\cite{Sobol:2026nfh} identified a regime of strong backreaction admitting a stable (within homogeneous backreaction) equilibrium for \(\xi\), but only for values of the particle production parameter satisfying \(\xi \lesssim 3.1\). The solution identified in the present work is distinct, since it occurs in the mild backreaction regime and the corresponding stable solution satisfies \(\xi > 3.1\).} The solution itself, is however well-known: since the impact of axion gradients remain small and $\xi$ remains sizable and approximately constant for several e-folds, all conditions are met to recover the instantaneous backreaction Anber-Sorbo solution~\cite{Anber:2009ua}.

In addition to this feature, we find a marked difference between our two implementations of the criterion~\eqref{dB-B-01}. The first (GEF-based) implementation is triggered only in the strong backreaction regime, close to the onset of oscillations in the particle production parameter. Moreover, as in all previous examples, it is triggered near the point at which the gradient energy becomes comparable to the kinetic energy of the axion. The second (in-in-based) implementation, by contrast, is triggered much earlier, close to the moment at which the homogeneous backreaction trajectory first departs from the single-field slow-roll solution. Due to the fact that, as we mentioned, the prolonged epoch of mild backreaction is characterized by a slow steady state evolution, the strong separation in e-folds between the moments in which the two implementations of the perturbativity criterion are triggered corresponds to only a relatively modest increase in the value of the particle production parameter, from $\xi \simeq 4.12$ to $\xi \simeq 5.05$. 

The different impact on the axion trajectory can be understood quantitatively by comparing the gauge-field backreaction term with the Hubble friction. For $\beta=50$, the ratio ${\left|(\beta/M_p)\mathcal{B}^{(0)}\right|}/{
\left|3H\dot{\phi}\right|}$ is approximately $1.2\%$ when the in-in criterion is first triggered, at $N\simeq25$. Thus, a $10\%$ correction to $\mathcal{B}^{(0)}$ corresponds to a change in the axion term of only approximately $0.12\%$ of Hubble friction, explaining its small immediate effect on the trajectory. By contrast, when the GEF criterion is triggered, at $N\simeq58$, the gauge-field backreaction term is approximately $80\%$ of
Hubble friction, so modifications to this term have a much larger impact on the axion evolution.

Beyond this difference in the relative importance of the backreaction term, an inspection of the backreaction spectral shape offers further insight into the separation between the two crossing times. We note that for smaller couplings, the interference term (the brown line in Fig~\ref{fig:spec-15}, see also Figs.~\ref{fig:spec-18}--\ref{fig:spec-25} in App.~\ref{app:spectra}) shows a (IR to UV) redistribution of the power of the gauge modes. This gives rise to a significant change of the backreaction term, such that, as soon as the amplitude of the interference term becomes significant (triggering the in-in based implementation), one witnesses an immediate modification of the axion trajectory (triggering the GEF-based implementation). In a sense, the interference term might be underestimating the full importance of the contribution sourced by the axion gradients, since, as we already remarked, it does not have support for momenta greater than those of the `unperturbed' modes. Therefore, even a ${\rm O} \left( 10\% \right)$ contribution of the interference diagram (as needed to trigger the perturbativity criterion) in this range of momenta can signify a more significant effect from the sourced-sourced contribution, if the latter appears to be supported at larger momenta (that cannot be directly probed by the interference term). On the contrary, the spectral form of the interference term in Fig.~\ref{fig:spec-50} tracks the one of the `unperturbed' spectrum. This suggest that the IR to UV redistribution of power has still to take place, so that a ${\rm O} \left( 10\% \right)$ level of the interference diagram would be associated with a non-negligible but still subdominant overall contribution from the modes modified by the axion gradients. This would explain why the first implementation of the criterion, more closely related to the evolution and self-consistently tracking the impact of the modified backreaction term, triggers only at a significantly later time.

In light of this discussion, one might regard the in-in implementation of the criterion as sufficient, though possibly not necessary, for perturbativity. As long as it is satisfied, it guarantees that the system is evolving according to homogeneous backreaction. The strong $\beta = 50$ coupling that we have studied shows an example in which homogeneous backreaction continues to offer an adequate description of the dynamics, with the axion gradients providing a sizeable but yet not dominant contribution. Extending the in-in formulation of the criterion by either (i) making use of the spectral information obtained at one loop, or (ii) evaluating the two-loop $\left\langle A^{(s)} \, A^{(s)} \right\rangle$ contribution, or, ultimately, performing lattice simulations for these larger couplings will shed further light on this issue. 
%
\subsection{Results at constant $\xi$ \& $H$}
\label{sec:constant}
%
\begin{figure}
    \centering
    \includegraphics[width=0.66\linewidth]{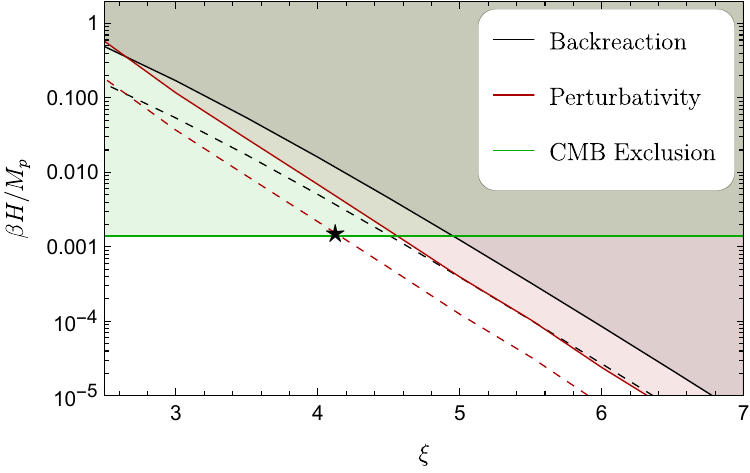}
    \caption{Results for constant $\xi$ and $H$. Parameters above the solid (respectively, dashed) black line indicate strong (respectively, mild) backreaction from the gauge fields, namely $\sigma > 1$ (respectively, $\sigma > 10^{-1}$) in Eq.~\eqref{eq:backreaction}. Parameters above the solid (respectively, dashed) red line indicate a violation of perturbativity, with the solid line corresponding to an equality in Eq.~\eqref{eq:perturbativity} and the dashed line to a suppression of the one-loop contribution by a factor 10. 
    Parameters above the green horizontal line are incompatible with CMB normalization if the axion is the inflaton (this follows from requiring that $H$ and $\xi$ are approximately constant for a few e-folds across the CMB window and not necessarily all throughout inflation). The black star is the exact moment in phase space at which the in-in implementation of the perturbativity criterion~\eqref{dB-B-01} is violated in the $\beta=50$ example analyzed in Sec.~\ref{sec:strong-coupling}.
    }
    \label{fig:constant-xi}
\end{figure}

To aid a systematic exploration of the parameter space, we can introduce a number of simplifying assumptions to the formalism underlying the in-in formulation of the criterion, with the aim of deriving simple analytic expressions that can be readily applied within the regime of validity of the approximation. Firstly, we assume that both the particle production parameter and the Hubble scale remain approximately constant over a time interval of ${\cal O}(\xi)$, so that the analytical results for the gauge mode functions can be used~\cite{Domcke:2020zez,Peloso:2022ovc}. Secondly, in order to facilitate comparison with previous work, we consider the ratio between the one-loop correction to the gauge-field power spectrum and the corresponding linear contribution, evaluated at horizon crossing,
\begin{eqnarray}
    \left\vert \; \frac{\delta^{(1)}\left\langle\hat{A}\left(\tau,\vec{k}\right)\hat{A}\left(\tau,-\vec{k}\right)\right\rangle}{\left\langle\hat{A}\left(\tau,\vec{k}\right)\hat{A}\left(\tau,-\vec{k}\right)\right\rangle} \; \right\vert_{k = a H} &\ll& 1 \;\;\;\;\; \left( {\rm Perturbativity \; condition} \right) \;.  \label{eq:perturbativity}
\end{eqnarray}

The criterion in Eq.~\eqref{eq:perturbativity} differs from that used in the preceding sections in two respects: it concerns the gauge-field two-point function rather than the backreaction, and it is evaluated at horizon crossing rather than integrated over momentum. The first difference is small within the constant-$\xi$, constant-$H$ amplified-mode approximation: the backreaction spectrum is proportional to $k\,\partial_\tau |A|^2$, and the relation $A'\simeq\sqrt{2\xi k/(-\tau)}\,A$ gives approximately the same multiplicative factor in the corrected and unperturbed spectra. Consequently, their relative corrections approximately coincide mode by mode.

The second difference remains relevant. The integrated
backreaction ratio is a weighted average of the spectral correction ratio, with the unperturbed backreaction spectrum providing the weight. Our choice of $k=aH$ is motivated by the concentration of the backreaction in a band of modes around the horizon scale, as illustrated by the spectral plots in the preceding sections: horizon crossing therefore provides a representative reference scale for this diagnostic, rather than an exact replacement for momentum integration. Its adequacy is supported by Fig.~\ref{fig:constant-xi}, where the black star, obtained using the momentum-integrated backreaction criterion in the slowly evolving $\beta=50$ example, lies close to the red dashed line obtained using the horizon-crossing two-point-function criterion at the same $10\%$ threshold. This agreement provides a numerical cross-check in that benchmark, without establishing exact equivalence of the two prescriptions throughout parameter space.

Under these assumptions, the computation reduces to an analysis closely analogous to that carried out in Ref.~\cite{Peloso:2016gqs}. The study of Ref.~\cite{Peloso:2016gqs} focused on CMB scales and therefore fixed the vacuum scalar power spectrum, \(P_{\zeta,v}\equiv H^4/(4\pi^2\dot{\phi}^2)\), to its observed value. Imposing this choice introduces an additional constraint into the criterion, so that the final result can be expressed as a function of the single parameter \(\xi\). Under this assumption, the onset of strong violation of perturbativity was found to occur for \(\xi \gtrsim 4.4\).\footnote{This is the value quoted in Fig.~9 of Ref.~\cite{Peloso:2016gqs}, where the Whittaker-function solution for the linear gauge-field mode functions was used. A slightly different threshold, \(\xi \gtrsim 4.6\), is quoted in the main text, Eq.~(5.10), where instead a power-law approximation for the mode functions was employed.}

Here we generalize this result to arbitrary values of \(P_{\zeta,v}\), and compute the corresponding conditions for both backreaction and perturbativity using the exact Whittaker-function solutions for the gauge-field mode functions while performing the relevant integrals numerically. We maintain only the gauge field helicity that is tachyonically enhanced and disregard the one that remains stable. Conveniently, both conditions can be expressed in the \((\xi,\beta H/M_p)\) plane. We display them in Fig.~\ref{fig:constant-xi}, where parameters above the black line are associated with significant backreaction from the gauge field (specifically, the solid line indicates strong backreaction, namely $\sigma = 1$ in Eq.~\eqref{eq:backreaction}, while the dashed line indicates mild backreaction, namely $\sigma = 10^{-1}$ in that equation), while parameters above the solid (respectively, dashed) red line violate perturbativity according to Eq.~\eqref{eq:perturbativity}. We observe that the red shaded region fully encloses the black one for $\xi \gtrsim 2.5$ or, equivalently, $\beta H/M_p \lesssim 1$,  implying that the strong backreaction regime cannot be reached without simultaneously violating perturbativity under the assumption of nearly constant $\xi$ and $H$. The lower limit of the horizontal axis of Fig.~\ref{fig:constant-xi} has been strategically chosen to be $\xi=2.5$ because that is the lowest value of the particle production parameter for which the calculation can be carried out reliably without a more sophisticated regularization and renormalization approach \cite{Ballardini:2019rqh,Animali:2022lig}. For smaller values of the particle production parameter there is no clear cut scale separation between the vacuum modes and the tachyonically enhanced ones and therefore the explicit choice of UV cutoff would lead to ${\cal O}(1)$ uncertainties to the final result \cite{Durrer:2024ibi}.

In addition, we superimpose on this plane a black star indicating the point at which the in-in implementation of the perturbativity criterion is triggered in the \(\beta=50\) example discussed in Sec.~\ref{sec:strong-coupling} (where, we recall, $\xi$ and $H$ are dynamically evolving). This provides an important cross-check, since in that example the particle production parameter varies sufficiently slowly that the approximation of constant \(H\) and \(\xi\) should remain reasonably accurate. The fact that the black star lies close to the red dashed line corresponding to the mild perturbativity condition indicates that the calculation is self-consistent, and that the full numerical results are compatible with the approximate analytic treatment in regimes where \(H\) and \(\xi\) vary only mildly.

Finally, we have produced a semi-analytic expression for the perturbative limit shown as the red line in Fig.~\ref{fig:constant-xi}. The derivation of our expression is explained in detail in App.~\ref{app:perturbativity-analytic} and takes the form
\begin{eqnarray}
    \frac{\beta H}{M_p} \, = \, 5.6\cdot 10^2 \; \xi \; {\rm e}^{-\pi\xi},\;\;\;\;\;\; {\rm for}\;\;\;\;\;\; \xi\gtrsim 2.5,\;\;\;\;\;\;({\rm  Perturbativity \;threshold})\;.
\label{perturbativity-constantHxi}
\end{eqnarray}
One can easily check that upon fixing $P_{\zeta,v}\equiv H^4/(4\pi^2\dot{\phi}^2)=2.1\cdot 10^{-9}$ this expression reduces to $\xi = 4.4$, fully compatible with the results of Ref.~\cite{Peloso:2016gqs}. As noted above, this expression is only valid while the particle production is sufficiently large ($\xi\gtrsim 2.5$) and the scale separation between the vacuum contribution and the tachyonically enhanced modes is large enough. An extension of this result to smaller values of the particle production parameter would require including the non-tachyonically enhanced helicity as well as an improved method for regularizing and renormalizing the UV modes. Moreover, the expression in Eq.~\eqref{perturbativity-constantHxi} is within about ${\cal O}(10 \%)$ of precision with respect to the fully numerical result shown in the red line of Fig.~\ref{fig:constant-xi} in the interval $6 \gtrsim\xi\gtrsim 2.5$ with the precision decreasing to about ${\cal O}(20\%)$ up to values between $7 \gtrsim\xi\gtrsim 6$.

We note that the newly found~\cite{Sobol:2026nfh} regime of constant $H$ and $\xi$,  characterized by a significant and stable homogeneous backreaction, is well above the line~\eqref{perturbativity-constantHxi}, therefore failing to meet the criterion of perturbativity. The last studied example  ($\beta = 50$) of the last section appears to suggest that this is a sufficient, but possibly not necessary condition for perturbativity. Similarly to what done in this example, a more definite answer on the stability (against axion gradients) within the regime found in~\cite{Sobol:2026nfh} might require implementing this regime in a concrete model and dynamical context, and studying the correction to the spectral backreaction in the in-in formalism or the axion dynamics in a $\delta\phi$GEF (or lattice) computation. 

\subsection{A model independent upper bound on $\beta \, H/M_p$ from CMB normalization}

We conclude the main part of this work by deriving an upper limit on the combination $\beta H/M_p$ for inflationary trajectories that seek to remain compatible with CMB observations, in a way that is as general and as independent of the inflationary potential as possible. Our starting point is the requirement that, irrespective of the form of the potential, the particle production parameter at CMB scales must satisfy~\cite{Barnaby:2011vw}
\begin{eqnarray}
    \xi_{\rm CMB}\lesssim 2.5 \;\;\;\;\; ({\rm small\;scalar\;non\mbox{-}Gaussianity})\;.
\end{eqnarray}
This condition, which is valid under the assumption that $\xi$ and $H$ remain approximately constant over the CMB window, implies that both the scalar and tensor power spectra are vacuum dominated, so that the sourced contributions are negligible at CMB scales. One may then relate the particle production parameter at CMB scales to the Hubble rate at the same epoch. Using $\epsilon_\varphi\equiv \dot{\phi}^2/(2H^2M_p^2)$, we obtain
\begin{align}
    \xi_{\rm CMB}\equiv \frac{\beta\dot{\phi}_{\rm CMB}}{2 M_p H_{\rm CMB}}
    = \frac{\beta\sqrt{\epsilon_{\varphi,{\rm CMB}}}}{\sqrt{2}}
    \overset{(\epsilon_\varphi\simeq\epsilon_H)}{\simeq}
    \frac{\beta\sqrt{\epsilon_{H,{\rm CMB}}}}{\sqrt{2}}
    \overset{(r\simeq 16\epsilon_H)}{\simeq}
    \frac{\beta\sqrt{r_{\rm CMB}}}{4\sqrt{2}}
    \simeq 1740\,\frac{\beta H_{\rm CMB}}{M_p}\;,
    \label{eq:theorem}
\end{align}
where we used the one-to-one relation between the tensor-to-scalar ratio and the Hubble scale at CMB scales, $r_{\rm CMB}\simeq 9.65\cdot 10^7\left(H_{\rm CMB}/M_p\right)^2$ in the last approximate equality.

Given the upper bound on $\xi$ at CMB scales imposed by non-Gaussianity constraints, Eq.~\eqref{eq:theorem} immediately translates this into an upper bound on the combination $\beta H_{\rm CMB}/M_p$. Since the Hubble rate is a monotonically decreasing function of time in the class of models considered here, this constraint can be propagated to all scales smaller than the CMB scales, and therefore to all subsequent times during inflation after the CMB modes exit the horizon. Therefore\footnote{We verified that all the examples studied in Refs.~\cite{Barbon:2025wjl,Franciolini:2026cps} satisfy this bound. It is moreover consistent with the results of the parameter space study in \cite{Sobol:2026nfh}.}
\begin{equation}
{\rm Axion \; inflation \; with } \; H ,\, \xi \simeq {\rm constant \; over \; CMB \; window} \;\; \Rightarrow \;\;  \frac{\beta \, H}{M_p} \lesssim 1.4 \cdot 10^{-3} \;. 
\label{beH-bound} 
\end{equation} 
This excludes the entire region above the green horizontal line in Fig.~\ref{fig:constant-xi}. This is the reason why we do not pursue tests of our criteria at still larger couplings or for alternative potentials. Our choice $\beta=50$ already saturates this upper bound. An additional implication of this theorem is that it is highly unlikely that one can realize a trajectory with strong backreaction while simultaneously maintaining $\xi\lesssim 3.1$, as in the stable solutions found in~\cite{Sobol:2026nfh}, and still remain compatible with the non-Gaussianity bounds at CMB scales. This conclusion can be circumvented in multi field models in which the axion is not the source of the observed CMB perturbations such as in the axion spectator scenario \cite{Peloso:2016gqs}. Other potential means to circumnavigate this constraint is to consider scenarios where the effective axion decay constant changes with time \cite{Linde:1991km} or by considering higher order operators or non-minimal coupling to gravity \cite{Almeida:2020kaq}.
%
\section{Conclusions}
\label{sec:conclusions}
%
In this work we have studied axion inflation coupled to Abelian gauge fields with a particular focus on the strong-backreaction regime, where gauge-field production due to the motion of the axion substantially modifies its dynamics. Several works in the literature have studied this system in the so-called homogeneous backreaction regime, in which the axion inhomogeneities are neglected (while the gauge field depends both on time and position). Our central goal has been to determine when axion inhomogeneities become important enough to invalidate this treatment and to identify the regime in which nonlinear effects play a relevant role. We improved in two distinct ways over existing works that studied the validity of this regime in the case of constant Hubble rate $H$ and gauge production parameter $\xi$ by studying the modifications to the gauge field and axion $2-$point correlation functions; firstly, we considered more realistic dynamical background solutions; secondly, since the backreaction on the axion zero mode is controlled by the quantity \(\langle \vec{E}\cdot\vec{B}\rangle\), we have used this observable as the organizing principle of our analysis and developed perturbative criteria that quantify how this correlator is modified in the presence of axion fluctuations. We introduced two complementary procedures to perform this evaluation. The first one is based on the first-order extension of the gradient expansion formalism, which perturbatively incorporates axion gradients into the dynamical evolution of the gauge-field correlators. The second one is based on a one-loop in-in computation of the correction to \(\langle \vec{E}\cdot\vec{B}\rangle\) that uses homogeneous backreaction as the `unperturbed' starting point, and the coupling between the gauge fields and the axion inhomogeneities as the interaction vertex. In both approaches, we take the onset of relevant nonlinear effects to occur when the corrected backreaction differs from the homogeneous result at the \(10\%\) level.

Using a quadratic potential as a benchmark, we first compared the outcome of these computations with the limited set of cases for which lattice simulations are currently available. In this regime, the two procedures are in broad agreement with one another and compatible with the lattice results. In particular, they indicate that axion inhomogeneities begin to substantially affect the backreaction once the axion gradient energy reaches the few percent level relative to the axion kinetic energy. This provides nontrivial support for both procedures and suggests that they capture the onset of the same underlying physical effect.

Extending the analysis to larger couplings reveals a richer picture. For moderately larger couplings, the two procedures remain broadly consistent. For sufficiently large couplings, however, our diagnostic tools point to a particularly interesting example, characterized by a prolonged stage of steady state evolution, in which the production parameter $\xi$ increases gradually and monotonically, and in which (i) $\xi$ is sufficiently large as to lead to observable signatures and (ii) the friction on the axion motion due to the gauge field production is subdominant but not-negligible with respect to the standard Hubble friction term. We term this regime as {\it mild backreaction}. To our knowledge, this is the first example in which the instantaneous backreaction Anber-Sorbo solution is dynamically realized in a phenomenologically interesting regime, since the examples from the previous literature are either characterized by a too small value of $\xi$ to lead to observable signatures, or by fast oscillations of $\xi$, once the backreaction from the gauge field becomes significant. While this solution was obtained in a toy-model which does not respect the latest CMB constraints, it provides an important proof of principle and indications on the regimes of parameter space which deserve further phenomenological study.

This example is characterized by a  clear separation between the two implementations of the perturbativity criterion in the strong-coupling regime. While the GEF-based criterion is only violated once the system enters the strong backreaction regime and when the axion gradient energy becomes comparable to its kinetic energy, the in-in implementation signals a breakdown much earlier, at the point where the homogeneous backreaction solution first departs from single-field slow-roll evolution. Although we are unable to fully explain the origin of this difference, the spectral shape of the one-loop backreaction term suggest a possible explanation. In the previously studied cases, the one-loop term exhibited a clear IR-to-UV redistribution of power, so that the onset of a sizeable interference contribution coincided with an immediate modification of the background evolution. By contrast, in the strong-coupling case the interference spectrum closely follows the unperturbed one, indicating that this redistribution has not yet occurred. This implies that the in-in implementation may be sensitive to the earliest effects of axion gradients, while the GEF-based implementation only becomes relevant once these effects have grown sufficiently to alter the background dynamics. Consequently, the in-in implementation may be viewed as a sufficient, though not necessarily necessary, condition for the validity of the homogeneous backreaction description. The strong coupling case that we have studied illustrates that homogeneous backreaction can remain accurate even when axion gradients provide a significant, yet still subdominant, contribution. Our results call for further investigation of this regime, e.g.\ by  higher-order calculations and, ultimately, lattice simulations in the strong-coupling regime.

We have also shown that the in-in implementation admits a useful analytic treatment in the regime where \(H\) and \(\xi\) evolve adiabatically. Within the constant-\(\xi\), constant-\(H\) approximation, we derived simple perturbativity and backreaction bounds in the \((\xi,\beta H/M_p)\) plane and found that, under the assumption that these parameters remain (nearly) constant, the strong-backreaction region is generically contained within the region where perturbation theory is lost. 

Finally, under the very reasonable assumptions that the vacuum scalar modes dominate over the sourced ones at CMB scales (so to guarantee small non-Gaussianity) and of adiabatically varying \(H\) and \(\xi\) at CMB scales (as it is generally expected due to the observed scale invariance of the large-scale primordial perturbations), and making no assumption on their evolution in the later stages of inflation, combined with the upper bound on \(\xi_{\rm CMB}\) from scalar non-Gaussianity, we obtained a general constraint on the product between the gauge-axion coupling and the Hubble rate, $\beta \, H / M_p \lesssim 1.4 \cdot 10^{-3}$, for all inflation models described by Eq.~\eqref{action}. If this bound is violated, the axion cannot be simultaneously the inflaton and the source of the primordial density perturbation, but some additional dynamical field(s) must be included.

Overall, our results provide a practical framework for assessing the regime of validity of perturbative treatments in axion inflation and for identifying the regions of parameter space where genuinely nonlinear dynamics must be taken into account. While we have focused our computation to the case of a quadratic inflaton potential, and probed how our criterion behaves at different axion-gauge couplings, our computations can be readily implemented for generic inflaton potentials. We expect that they can serve both as phenomenological diagnostics and as useful guides for future lattice studies, helping to focus numerical efforts on the most relevant and most challenging regimes of axion inflation.

\begin{acknowledgments}
We thank Kai Schmitz, Sasha Sobol and Lorenzo Sorbo for insightful discussions. We are particularly grateful to Kai Schmitz for very helpful comments on the draft.
We acknowledge support from the DOE Topical Collaboration “Nuclear Theory for New Physics” award No. DE-SC0023663.
A.P. acknowledges the “Consolidación Investigadora” grant CNS2022-135590. The work of A.P. is partially supported by the Spanish Research Agency (Agencia Estatal de Investigación) through the Grant IFT Centro de Excelencia Severo Ochoa No CEX2020-001007-S, funded by MCIN/AEI/10.13039/501100011033. M.P. acknowledges support from Istituto Nazionale di Fisica Nucleare (INFN) through the Theoretical Astroparticle Physics (TAsP) project. 
S.S. was supported by the U.S. Department of Energy Office and by the Laboratory Directed Research and Development (LDRD) program of Los Alamos National Laboratory under project numbers 20230047DR, 20250164ER and 20260246ER. Los Alamos National Laboratory is operated by Triad National Security, LLC, for the National Nuclear Security Administration of the U.S. Department of Energy (Contract No. 89233218CNA000001)
\end{acknowledgments}

\appendix

\section{Details of the loop in-in computation}
\label{app:loop}

\BeforeBeginEnvironment{equation}{\begingroup\small}
\AfterEndEnvironment{equation}{\endgroup}

\BeforeBeginEnvironment{eqnarray}{\begingroup\small}
\AfterEndEnvironment{eqnarray}{\endgroup}

In this appendix we provide some intermediate steps in the evaluation of the correlator \eqref{correlator-C0}. Inserting the decomposition \eqref{eq:A-deco} in the interaction Hamiltonian \eqref{Hint}, the inner commutator rewrites 
\begin{align} 
\begin{split}
\left[ A \left( \tau ,\, \vec{k}_1 \right) A \left( \tau' ,\, \vec{k}_2 \right) ,\, H_{\rm int} \left( \tau_1 \right) \right] = \frac{\beta}{M_p} \int& \frac{d^3 k d^3 p}{\left( 2 \pi \right)^{3/2}} \left\vert \vec{k} + \vec{p} \right\vert \epsilon_i \left( \vec{p} \right) \epsilon_i \left( - \vec{k} - \vec{p} \right) \delta {\hat \phi} \left( \tau_1 ,\, \vec{k} \right) \\
& 
\left[ {\hat A} \left( \tau ,\, \vec{k}_1 \right) {\hat A} \left( \tau' ,\, \vec{k}_2 \right) ,\, {\hat A}' \left( \tau_1 ,\, \vec{p} \right) {\hat A} \left( \tau_1 ,\, - \vec{k} - \vec{p} \right) \right] \,. 
\end{split}
\end{align} 

The commutator in the second line has the structure 
\begin{equation}
\left[ A_1 A_2 ,\, A_3 A_4 \right] = A_1 \left[ A_2,\, A_3 \right] A_4 + A_1 A_3 \left[ A_2 ,\, A_4 \right] + \left[ A_1 ,\, A_3 \right] A_4 \, A_2 + A_3 \left[ A_1 ,\, A_4 \right] A_2 \;,
\label{A4}
\end{equation} 
and we decompose 
\begin{eqnarray} 
\left[ A \left( \tau ,\, \vec{k}_1 \right) A \left( \tau' ,\, \vec{k}_2 \right) ,\, H_{\rm int} \left( \tau_1 \right) \right] = \sum_{i=1}^4 \left[ A \left( \tau ,\, \vec{k}_1 \right) A \left( \tau' ,\, \vec{k}_2 \right) ,\, H_{\rm int} \left( \tau_1 \right) \right]_i \;, 
\label{[]i}
\end{eqnarray} 
where the $i-$th term of this sum receives contribution from the $i-th$ term in \eqref{A4}. In the following, for brevity we discuss the evaluation of only the first of these terms. The other three ones are computed in an identical manner. The contribution of the first term to the outer commutator reads 
\begin{eqnarray} 
&& \!\!\!\!\!\!\!\!
\!\!\!\!\!\!\!\! 
\left\langle \left[ \left[ A \left( \tau ,\, \vec{k}_1 \right) A \left( \tau' ,\, \vec{k}_2 \right) ,\, H_{\rm int} \left( \tau_1 \right) \right]_1 ,\, H_{\rm int} \left( \tau_2 \right) \right] \right\rangle \nonumber\\
&& \quad\quad\quad\quad = \frac{2  i  \beta {\rm Im } \left[ A \left( \tau' ,\, k_2 \right) A^{'*} \left( \tau_1 ,\, k_2 \right) \right]}{M_p} \int \frac{d^3 k}{\left( 2 \pi \right)^{3/2}} \left\vert \vec{k} - \vec{k}_2 \right\vert \epsilon_i \left( -\vec{k}_2 \right) \epsilon_i \left( - \vec{k} + \vec{k}_2 \right) \nonumber\\ 
&& \quad\quad\quad\quad \Bigg\langle \Bigg[ \delta {\hat \phi} \left( \tau_1 ,\, \vec{k} \right)  {\hat A} \left( \tau ,\, \vec{k}_1 \right)   {\hat A} \left( \tau_1 ,\, - \vec{k} + \vec{k}_2 \right) ,\, \frac{\beta}{M_p} \int \frac{d^3 k' d^3 p'}{\left( 2 \pi \right)^{3/2}} \left\vert \vec{k}' + \vec{p}' \right\vert \epsilon_j \left( \vec{p}' \right) \epsilon_j \left( - \vec{k}' - \vec{p}' \right) \nonumber\\ 
&& \quad\quad\quad\quad \quad\quad\quad\quad \quad\quad\quad\quad \quad\quad\quad\quad  
\quad\quad\quad\quad  
\quad\quad\quad\quad  
\delta {\hat \phi} \left( \tau_2 ,\, \vec{k}' \right) {\hat A}' \left( \tau_2 ,\, \vec{p}' \right) {\hat A} \left( \tau_2 ,\, - \vec{k}' - \vec{p}' \right) \Bigg] \Bigg\rangle \;. \nonumber\\ 
\end{eqnarray} 

This expression has a contribution with structure $\left[ \delta \phi ,\, \delta \phi \right] \, \left\langle A^4 \right\rangle$ and one with structure $\left[ A ,\, A \right] \,\left\langle \delta \phi^2 A^2 \right\rangle$. We recall that the mode functions entering in these expressions are `unperturbed' ones. Namely, they are the solutions of Eqs. \eqref{eq:eom-dphi}. As these equations are linear in the perturbations, the `unperturbed' modes are Gaussian, and the the $\left\langle A^4 \right\rangle$ correlator can be expressed as the combinatorial sum of products of two-point functions, $\left\langle A_1 \, A_2 \, A_3 \, A_4 \right\rangle = \left\langle A_1 \, A_2 \right\rangle \left\langle A_3 \, A_4 \right\rangle +$ two permutations. The gauge field mode functions are enhanced by the pseudoscalar interaction, while the scalar field mode functions are not enhanced. The enhancement of the gauge mode functions is canceled in their commutator, namely, $\left\langle A \, A \right\rangle \gg \left[  A ,\, A \right]$. Therefore, the contribution proportional to $\left[ \delta \phi ,\, \delta \phi \right] \, \left\langle A^4 \right\rangle$ is subdominant and can be ignored~\cite{Peloso:2016gqs}. This is typically true for the in-in computation with highly amplified source, that becomes classical. Therefore
\begin{eqnarray} 
&& \!\!\!\!\!\!\!\!  \left\langle \left[ \left[ A \left( \tau ,\, \vec{k}_1 \right) A \left( \tau' ,\, \vec{k}_2 \right) ,\, H_{\rm int} \left( \tau_1 \right) \right]_1 ,\, H_{\rm int} \left( \tau_2 \right) \right] \right\rangle \simeq - \frac{4 \beta^2}{M_p^2} 
\int \frac{d^3 k d^3 p}{\left( 2 \pi \right)^3}  \left\vert \vec{k} - \vec{k}_2 \right\vert \left\vert - \vec{k} + \vec{p} \right\vert \nonumber\\
&& \quad\quad\quad\quad
{\rm Im } \left[ A \left( \tau' ,\, k_2 \right) A^{'*} \left( \tau_1 ,\, k_2 \right) \right]
{\rm Im } \left[ \delta \phi \left( \tau_1 ,\, k \right) \delta \phi^* \left( \tau_2 ,\, k \right) \right] \epsilon_i \left( -\vec{k}_2 \right) \epsilon_i \left( - \vec{k} + \vec{k}_2 \right) \epsilon_j \left( \vec{p} \right) \epsilon_j \left( \vec{k} - \vec{p} \right)  \nonumber\\ 
&& \quad\quad\quad\quad 
\left\langle   {\hat A} \left( \tau ,\, \vec{k}_1 \right)   {\hat A} \left( \tau_1 ,\, - \vec{k} + \vec{k}_2 \right)    {\hat A}' \left( \tau_2 ,\, \vec{p} \right) {\hat A} \left( \tau_2 ,\, \vec{k} - \vec{p} \right) \right\rangle \;. 
\end{eqnarray} 

As mentioned, the $4-$point correlator can be decomposed as products of $2-$point correlators of commuting fields 
\begin{eqnarray}
\left\langle {\hat A} \left( \tau ,\, \vec{k} \right) \, {\hat A} \left( \tau' ,\, \vec{k}' \right) \right\rangle &\simeq& {\rm Re } \left[ A \left( \tau ,\, \vec{k} \right) A^* \left( \tau' ,\, \vec{k}' \right) \right] \delta^{(3)} \left( \vec{k} + \vec{k}' \right) \;. \nonumber\\ 
\end{eqnarray} 
Lengthy but straightforward algebra then leads to 
\begin{eqnarray} 
&& \!\!\!\!\!\!\!\!  \!\!\!\!\!\!\!\!  \!\!\!\!\!\!\!\!  \!\!\!\!\!\!\!\! \left\langle \left[ \left[ A \left( \tau ,\, \vec{k}_1 \right) A \left( \tau' ,\, \vec{k}_2 \right) ,\, H_{\rm int} \left( \tau_1 \right) \right]_1 ,\, H_{\rm int} \left( \tau_2 \right) \right] \right\rangle \simeq - \frac{4 \beta^2}{M_p^2} \delta^{(3)} \left( \vec{k}_1 + \vec{k}_2 \right)  \nonumber\\ 
&& \!\!\!\!\!\!\!\!  \!\!\!\!\!\!\!\!  \!\!\!\!\!\!\!\!  
\int \frac{d^3 p}{\left( 2 \pi \right)^3} \; \left\vert \epsilon_i \left( \vec{k}_1 \right) \epsilon_i \left(  - \vec{q} \right) \right\vert^2 \; q \; {\rm Im } \left[ A \left( \tau' ,\, k_1 \right) A^{'*} \left( \tau_1 ,\, k_1 \right) \right] {\rm Im } \left[ \delta \phi \left( \tau_1 ,\, p \right) \delta \phi^* \left( \tau_2 ,\, p \right) \right]  \nonumber\\ 
&& \Bigg\{ 
q \; {\rm Re } \left[ A \left( \tau ,\, k_1 \right) A^{'*} \left( \tau_2 ,\, k_1 \right) \right] \; {\rm Re } \left[ A \left( \tau_1 ,\, q \right) A^* \left( \tau_2 ,\, q \right) \right] \nonumber\\ 
&&  \quad\quad + \; k_1 \; {\rm Re } \left[ A \left( \tau ,\, k_1 \right) A^* \left( \tau_2 ,\, k_1 \right) \right] \; {\rm Re } \left[ A \left( \tau_1 ,\, q \right) A^{'*} \left( \tau_2 ,\, q \right)  \right] \Bigg\} \Big\vert_{\vec{q} \equiv \vec{k}_1 - \vec{p}} \;. 
\label{[]1}
\end{eqnarray} 

The polarization operators satisfy (we recall that we are considering only the $\lambda = +$ enhanced gauge polarization) 
\begin{equation}
\left\vert \epsilon_i \left( \vec{k}_1 \right) \epsilon_i \left( - \vec{q} \right) \right\vert^2 = \frac{\left( 1 + {\hat k}_1 \cdot {\hat q} \right)^2}{4} \;. 
\end{equation} 

We use this expression in \eqref{[]1}. We then evaluate the other three terms in \eqref{[]i} analogously, and we add the four contributions, to obtain 
\begin{eqnarray} 
&& \!\!\!\!\!\!\!\!  \!\!\!\!\!\!\!\!  \!\!\!\!\!\!\!\!  \!\!\!\!\!\!\!\!
{\cal C}^{(0,0)} \left( \tau ,\, \tau'  ,\, \tau_1 ,\, \tau_2 ,\, k_1 \right) \simeq - \frac{\beta^2}{M_p^2} \int \frac{d^3 p}{\left( 2 \pi \right)^3} \; \left( 1 + {\hat k}_1 \cdot {\hat q} \right)^2 {\rm Im } \left[ \delta \phi \left( \tau_1 ,\, p \right) \delta \phi^* \left( \tau_2 ,\, p \right) \right]    \nonumber\\ 
&& \!\!\!\!\!\!\!\!  
\Bigg\{ q \; {\rm Im } \left[ A \left( \tau' ,\, k_1 \right) A^{'*} \left( \tau_1 ,\, k_1 \right) \right]   \Bigg[ 
q \; {\rm Re } \left[ A \left( \tau ,\, k_1 \right) A^{'*} \left( \tau_2 ,\, k_1 \right) \right] \; {\rm Re } \left[ A \left( \tau_1 ,\, q \right) A^* \left( \tau_2 ,\, q \right) \right] \nonumber\\ 
&& \quad\quad\quad\quad  \quad\quad\quad\quad \quad\quad\quad\quad \quad 
 + \; k_1 \; {\rm Re } \left[ A \left( \tau ,\, k_1 \right) A^* \left( \tau_2 ,\, k_1 \right) \right] \; {\rm Re } \left[ A \left( \tau_1 ,\, q \right) A^{'*} \left( \tau_2 ,\, q \right)  \right] \Bigg] \nonumber\\ 
&& \!\!\!\!\!\!\!\!   
+ k_1 \;  {\rm Im } \left[ A \left( \tau' ,\, k_1 \right) A^* \left( \tau_1 ,\, k_1 \right) \right]  \Bigg[ q \; {\rm Re } \left[ A \left( \tau ,\, k_1 \right) A^{'*} \left( \tau_2 ,\, k_1 \right) \right] \; {\rm Re } \left[ A' \left( \tau_1 ,\, q \right) A^* \left( \tau_2 ,\, q \right) \right] \nonumber\\ 
&& \quad\quad\quad\quad  \quad\quad\quad\quad 
\quad\quad\quad\quad \quad +   k_1 \; {\rm Re } \left[ A \left( \tau ,\, k_1 \right) A^* \left( \tau_2 ,\, k_1 \right) \right] \; {\rm Re } \left[ A' \left( \tau_1 ,\, q \right) A^{'*} \left( \tau_2 ,\, q \right)  \right]  \Bigg] \nonumber\\ 
&& \!\!\!\!\!\!\!\!  
+ q \; {\rm Im } \left[ A \left( \tau ,\, k_1 \right) A^{'*} \left( \tau_1 ,\, k_1 \right) \right]    \Bigg[ q \; {\rm Re } \left[ A \left( \tau' ,\, k_1 \right) A^{'*} \left( \tau_2 ,\, k_1 \right)  \right]  \; {\rm Re } \left[ A \left( \tau_1 ,\, q \right) A^* \left( \tau_2 ,\, q \right) \right] \nonumber\\ 
&& \quad\quad\quad\quad  \quad\quad\quad\quad 
\quad\quad\quad\quad \quad +  k_1 \; {\rm Re } \left[ A \left( \tau' ,\, k_1  \right) A^* \left( \tau_2 ,\, k_1 \right) \right] \;{\rm Re } \left[ A \left( \tau_1 ,\, q \right) A^{'*} \left( \tau_2 ,\, q \right) \right] \Bigg] \nonumber\\ 
&& \!\!\!\!\!\!\!\!  
+ k_1 \;   {\rm Im } \left[ A \left( \tau ,\, k_1 \right) A^* \left( \tau_1 ,\, k_1 \right) \right]  \Bigg[ q \; {\rm Re } \left[ A \left( \tau' ,\, k_1 \right) A^{'*} \left( \tau_2 ,\, k_1 \right)  \right]   \; {\rm Re } \left[ A' \left( \tau_1 ,\, q \right) A^* \left( \tau_2 ,\, q \right) \right] \nonumber\\ 
&& \quad\quad\quad\quad  \quad\quad\quad\quad 
\quad\quad\quad\quad \quad + k_1 \; {\rm Re } \left[ A \left( \tau' ,\, k_1  \right) A^* \left( \tau_2 ,\, k_1 \right) \right]   \; {\rm Re } \left[ A' \left( \tau_1 ,\, q \right) A^{'*} \left( \tau_2 ,\, q \right) \right] \Bigg] \nonumber\\ 
&&   \Bigg\} \Big\vert_{\; \vec{q} \; \equiv \; \vec{k}_1 - \vec{p}} \;. 
\label{Atau-Ataup}
\end{eqnarray} 

expression is used in Eq.~\eqref{dB1-final}  of the main text to evaluate the one-loop correction to the gauge field backreaction on the evolution of the background axion. 

The numerical evaluation follows the strategy employed in Ref.~\cite{Garcia-Bellido:2023ser}. We insert Eq.~(A.9) into Eq.~(3.13) and interchange the internal momentum and time integrations, performing the time integrals first at fixed
internal momenta. For each external wave number $k$, we parametrize the internal momentum magnitudes as
\[
p \equiv |\vec{p}| = k(X+Y), \qquad q \equiv |\vec{k}-\vec{p}| = k(X-Y),
\]
and introduce the logarithmic variable $\Sigma \equiv \ln X$. We construct a $50\times50$ grid with uniformly spaced points in each of the variables $(\Sigma,Y)$, covering the ranges
\[
\ln(1/2) \leq \Sigma \leq \ln(1000),
\qquad
-\frac{1}{2} \leq Y \leq \frac{1}{2}.
\]

At each point of this momentum grid, we evaluate the nested time integrals using the two-dimensional trapezoidal rule, with a time step of $\Delta N=0.25$ e-folds. For an evaluation time $N$, the integration is restricted to the time-ordered domain $N-7 \leq N_2 \leq N_1 \leq N$, corresponding to a look-back interval of seven e-folds. This interval is sufficient to capture the relevant contributions and achieve convergence in the examples considered here. Once the time integrals have been evaluated at all momentum-grid points, we perform the remaining integration over $(\Sigma,Y)$, including the corresponding Jacobian, again using the two-dimensional trapezoidal rule. We find that this procedure yields convergent results for all the examples studied in this work.

To exclude ultraviolet vacuum contributions unrelated to gauge-field production, we impose a time-dependent momentum cutoff on all scalar and gauge-field mode functions entering the integrand. Specifically, a mode function evaluated at time $N_i$ is set to zero whenever its comoving wave number exceeds $k_{\rm max}(N_i)$, defined in Eq.~(4.5) as the largest wave number that has entered the gauge-field tachyonic band at any earlier time. Applying this prescription to both the internal and external mode functions restricts the calculation to momentum scales that have already been reached by the tachyonic instability and excludes contributions from the unamplified ultraviolet sector.

Turning our attention back to the analytic expression~\eqref{Atau-Ataup}, as a check, we want to show that this expression, when evaluated with the approximate analytical mode functions \eqref{A-constant-xiH}, reproduces the correction to the equal-time gauge field $2-$point function given in~\cite{Peloso:2016gqs}. To this purpose, we evaluate Eq. \eqref{Atau-Ataup} at equal external times (shifting the integration momentum to $\vec{p} \to \vec{k}_1 + \vec{p}$) and add the internal time integrations, to write
\begin{eqnarray} 
&& \!\!\!\!\!\!\!\! \!\!\!\!\!\!\!\! \!\!\!\!\!\!\!\! 
\delta^{(1)} \left\langle {\hat A} \left( \tau ,\, \vec{k}_1 \right) \, {\hat A} \left( \tau ,\, \vec{k}_2 \right) \right\rangle \simeq \frac{2 \beta^2}{M_p^2} \; \delta^{(3)} \left( \vec{k}_1 + \vec{k}_2 \right) \, \int^\tau d \tau_1 \int^{\tau_1} d \tau_2 \nonumber\\
&&   \int \frac{d^3 p}{\left( 2 \pi \right)^3} \left( 1 - {\hat k}_1 \cdot {\hat p} \right)^2 \; {\rm Im } \left[ \delta \phi \left( \tau_1 ,\, \left\vert \vec{k}_1 + \vec{p} \right\vert  \right) \delta \phi^* \left( \tau_2 ,\, \left\vert \vec{k}_1 + \vec{p} \right\vert \right) \right] \nonumber\\
&& \quad\quad\quad\quad   \Bigg\{  p^2 \, {\rm Im } \left[ A \left( \tau ,\, k_1 \right) A^{'*} \left( \tau_1 ,\, k_1 \right) \right] {\rm Re } \left[ A \left( \tau ,\, k_1 \right) A^{'*} \left( \tau_2 ,\, k_1 \right) \right]  {\rm Re } \left[ A \left( \tau_1 ,\, p \right) A^* \left( \tau_2 ,\, p \right) \right]  \nonumber\\ 
&& \quad\quad\quad\quad       + p \, k_1 \,  {\rm Im } \left[ A \left( \tau ,\, k_1 \right) A^{'*} \left( \tau_1 ,\, k_1 \right) \right] {\rm Re } \left[ A \left( \tau ,\, k_1 \right) A^* \left( \tau_2 ,\, k_1 \right) \right] {\rm Re } \left[ A \left( \tau_1 ,\,p \right) A^{'*} \left( \tau_2 ,\, p \right)  \right]  \nonumber\\ 
&& \quad\quad\quad\quad     +  k_1  \, p \, {\rm Im } \left[ A \left( \tau ,\, k_1 \right) A^* \left( \tau_1 ,\, k_1 \right) \right]  {\rm Re } \left[ A \left( \tau ,\, k_1 \right) A^{'*} \left( \tau_2 ,\, k_1 \right) \right] {\rm Re } \left[ A' \left( \tau_1 ,\, p \right) A^* \left( \tau_2 ,\, p \right) \right]  \nonumber\\ 
&& \quad\quad\quad\quad     + k_1^2 {\rm Im } \left[ A \left( \tau ,\, k_1 \right) A^* \left( \tau_1 ,\, k_1 \right) \right] {\rm Re } \left[ A \left( \tau ,\, k_1 \right) A^* \left( \tau_2 ,\, k_1 \right) \right] {\rm Re } \left[ A' \left( \tau_1 ,\, p \right) A^{'*}  \left( \tau_2 ,\, p \right) \right]  \Bigg\} \;. \nonumber\\ 
\label{d1AA-compa0}
\end{eqnarray} 

To compare with~\cite{Peloso:2016gqs} (where the functions \eqref{A-constant-xiH} are used in the evaluation of $\left\langle A^4 \right\rangle$), we disregard the imaginary part of the mode functions outside the gauge field commutator. Namely, 
\begin{align} 
\begin{split}
& p^2 \, {\rm Im } \left[ A \left( \tau ,\, k_1 \right) A^{'*} \left( \tau_1 ,\, k_1 \right) \right] {\rm Re } \left[ A \left( \tau ,\, k_1 \right) A^{'*} \left( \tau_2 ,\, k_1 \right) \right]  {\rm Re } \left[ A \left( \tau_1 ,\, p \right) A^* \left( \tau_2 ,\, p \right) \right] \\ 
& \;\;\;\; \to \;\; p^2 \, {\rm Im } \left[ A \left( \tau ,\, k_1 \right) A^{'*} \left( \tau_1 ,\, k_1 \right) \right] A_R \left( \tau ,\, k_1 \right) A_R' \left( \tau_2 ,\, k_1 \right) \, A_R \left( \tau_1 ,\, p \right) A_R \left( \tau_2 ,\, p \right) \;, 
\end{split}
\end{align} 
(where $A_R$ denotes the real part of $A$), and analogously for the last three lines of \eqref{d1AA-compa0}. Straightforward algebra shows that the resulting expression agrees with Eq. (C.5) of~\cite{Peloso:2016gqs}. 
\section{Analytic perturbativity limit}
\label{app:perturbativity-analytic}
%
In the context of the in-in computation, one-loop corrections overcome the unperturbed solution beyond the red line in Fig.~\ref{fig:constant-xi}, which corresponds to enforcing an equality in Eq.~\eqref{eq:perturbativity}. In the appendix we produce a semi-analytic expression that closely tracks the numerically derived results shown in Fig.~\ref{fig:constant-xi}. This analytic expression takes the form of a fitted function whose functional dependence on $\xi$ and $H$ can be predicted by analyzing the form of the integral in Eq.~(\ref{d1AA-compa0}) closely. 

In Sec.~\ref{sec:axion-inflation-review} in Eq.~\eqref{A-constant-xiH} we wrote only the real part of the solution to the gauge field equations of motion for constant $\xi$ and $H$ as that is the one that is exponentially enhanced. Instead a more complete solution that accounts for the suppressed imaginary part can be derived as a direct result of the Wronskian condition~\cite{Peloso:2016gqs} $A_+A_+^{'*}-c.c.=i$, namely
\begin{equation}
    A_+(\tau,k)\simeq \frac{1}{\sqrt{2k}}\left(\frac{-k\tau}{2\xi}\right)^{1/4}{\rm e}^{\pi\xi-2\sqrt{-2\xi k \tau}}+ \frac{i}{\sqrt{2k}}\left(\frac{-k\tau}{2^5\xi}\right)^{1/4}{\rm e}^{-\pi\xi+2\sqrt{-2\xi k \tau}}\;,
\label{A2-constant-xiH}
\end{equation}
This shows that the real and imaginary parts of the gauge mode functions feature an exponential dependence with respect to parameter $\xi$ with positive and negative exponents respectively. Furthermore, the derivatives of the gauge mode functions also share this feature
\begin{eqnarray}
    \frac{d A_+\left(\tau,k\right)}{d\tau} \simeq   \sqrt{\frac{k}{2}}\left(\frac{2\xi}{-k\tau}\right)^{1/4}{\rm e}^{\pi\xi-2\sqrt{-2\xi k \tau}}- i\sqrt{\frac{k}{2}}\left(\frac{\xi }{-8 k\tau}\right)^{1/4}{\rm e}^{-\pi\xi+2\sqrt{-2\xi k \tau}}\;,
\end{eqnarray}

With this in mind, we can predict the functional form of the red line in Fig.~\ref{fig:constant-xi} by schematically expanding Eq.~(\ref{d1AA-compa0}) in terms of the real and imaginary parts of the gauge field mode functions. We disregard the presence of derivatives acting on mode functions, since as we established above, the derivatives of the mode functions share the same exponential dependence as the mode functions themselves and we also suppress the time and momentum dependence for simplicity
\begin{eqnarray}
    \frac{\delta^{(1)}\left\langle AA\right\rangle}{\left\langle A A\right\rangle} \propto \frac{\beta^2 \;{\rm Im}[\delta\phi \delta\phi^*]\; {\rm Im}[AA^*] \;{\rm Re}[AA^*]\; {\rm Re}[AA^*]}{M_p^2\; {\rm Re}[A A^*]}\;.
\end{eqnarray}
Expanding now the mode functions in real and imaginary parts and preserving only the exponential dependence (i.e. ${\rm Re}[A]\propto {\rm e}^{+\pi\xi}$ and ${\rm Im}[A]\propto {\rm e}^{-\pi\xi}$), while also replacing $\delta\phi\propto H$,\footnote{The full expression for the scalar field mode functions is $\delta\phi(\tau,k)=\frac{H_k\left(1+i k \tau\right){\rm e}^{-ik\tau}}{\sqrt{2}k^{3/2}}$.} we obtain
\begin{eqnarray}
    \frac{\delta^{(1)}\left\langle AA\right\rangle}{\left\langle A A\right\rangle} \propto g(\xi)\;\frac{\beta^2 H^2 \;{\rm e}^{4\pi\xi}}{M_p^2\; {\rm e}^{2\pi\xi}}= g(\xi) \;\frac{\beta^2 H^2}{M_p^2} {\rm e}^{2\pi\xi}\;,
    \label{eq:fitting}
\end{eqnarray}
where $g(\xi)$ is some non-exponential function of $\xi$ to be determined by a fit to the numerical results. Upon inverting Eq.~\eqref{eq:fitting} and taking the square root on both sides we see that the red line must have the following parametric dependence
\begin{eqnarray}
    \frac{\beta H}{M_p}= g(\xi)^{-1/2}\;{\rm e}^{-\pi\xi}\;.
\end{eqnarray}
After performing a fit to the numerical results, with a polynomial $g(\xi)^{-1/2}$ we find excellent agreement for $g(\xi)^{-1/2} \,=\, 5.6\cdot 10^2\;\xi $. As a result the red line in Fig.~\ref{fig:constant-xi} is well approximated by
\begin{eqnarray}
    \frac{\beta H}{M_p} \, = \, 5.6\cdot 10^2 \; \xi \; {\rm e}^{-\pi\xi}
\end{eqnarray}
where the perturbative regime lies well below this line in Fig.~\ref{fig:constant-xi}.
%
\section{Additional backreaction spectra}
\label{app:spectra}
\begin{figure}[b]
    \centering
    \includegraphics[width=0.91\linewidth]{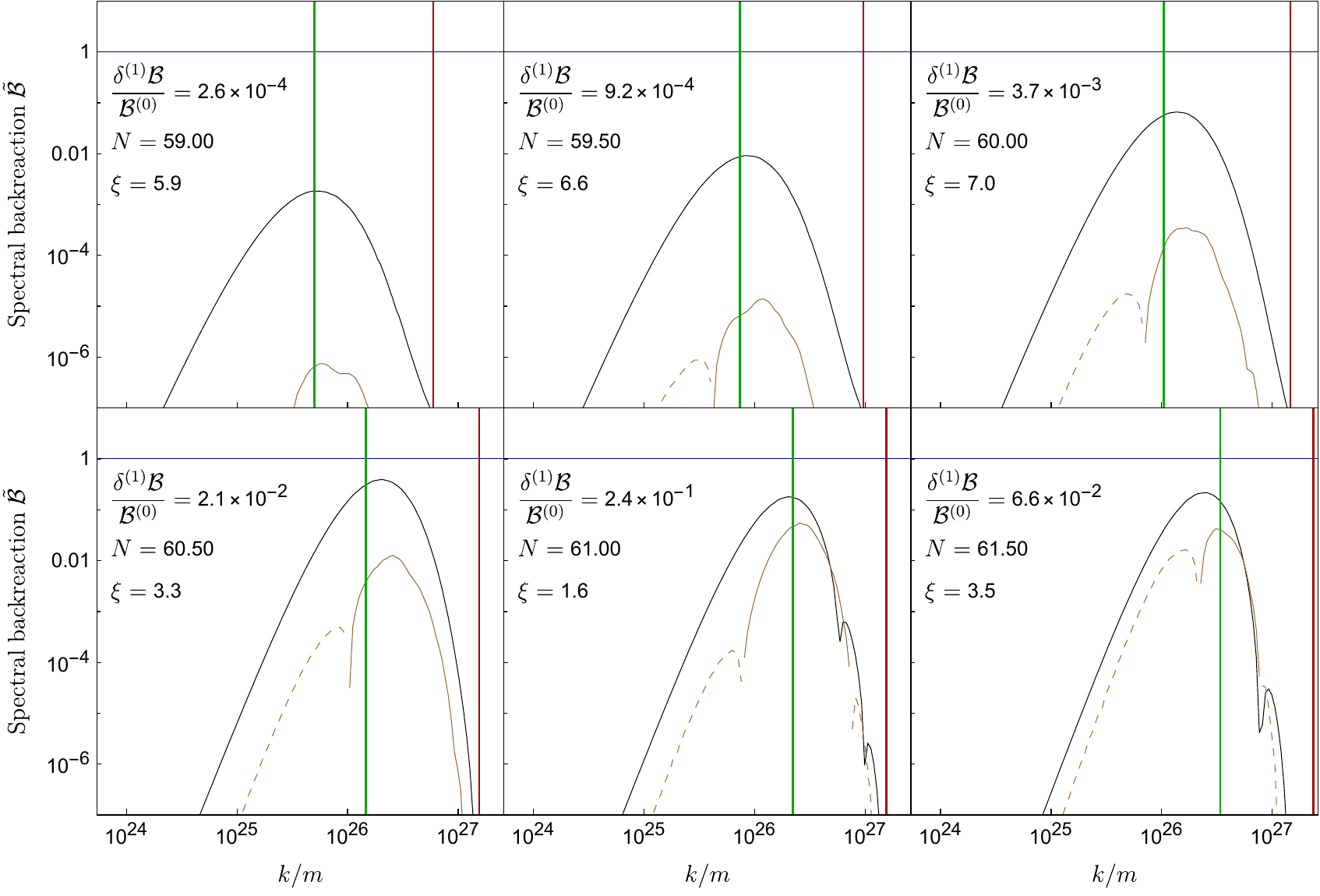}
    \caption{Normalized spectral backreaction for coupling $\beta=18$. Color coding as in Fig.~\ref{fig:spec-15}.
    }
    \label{fig:spec-18}
\end{figure}

\begin{figure}[h]
    \centering
    \includegraphics[width=0.91\linewidth]{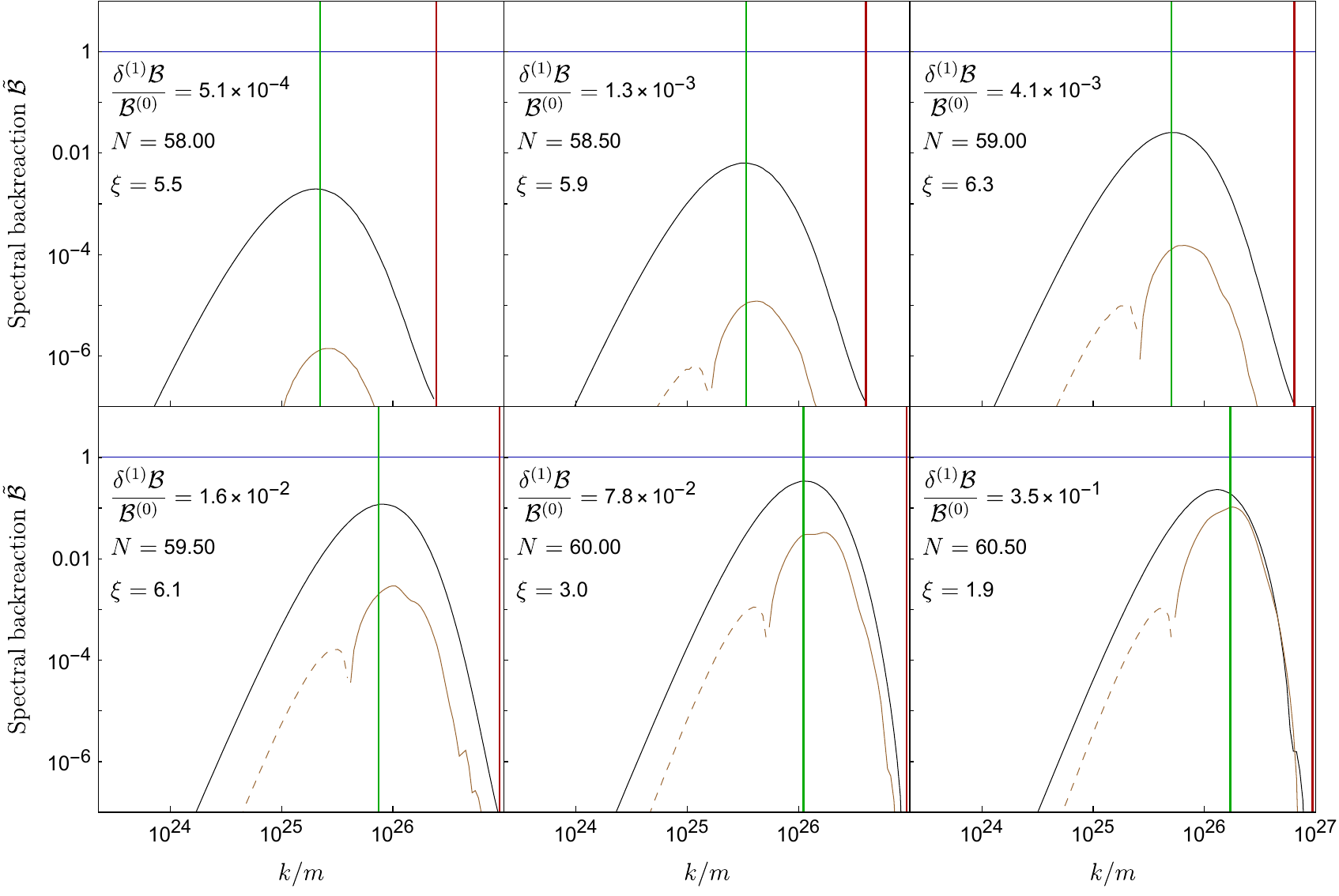}
    \caption{Normalized spectral backreaction for coupling $\beta=20$. Color coding as in Fig.~\ref{fig:spec-15}.}
    \label{fig:spec-20}
\end{figure}

\begin{figure}[b]
    \centering
    \includegraphics[width=0.91\linewidth]{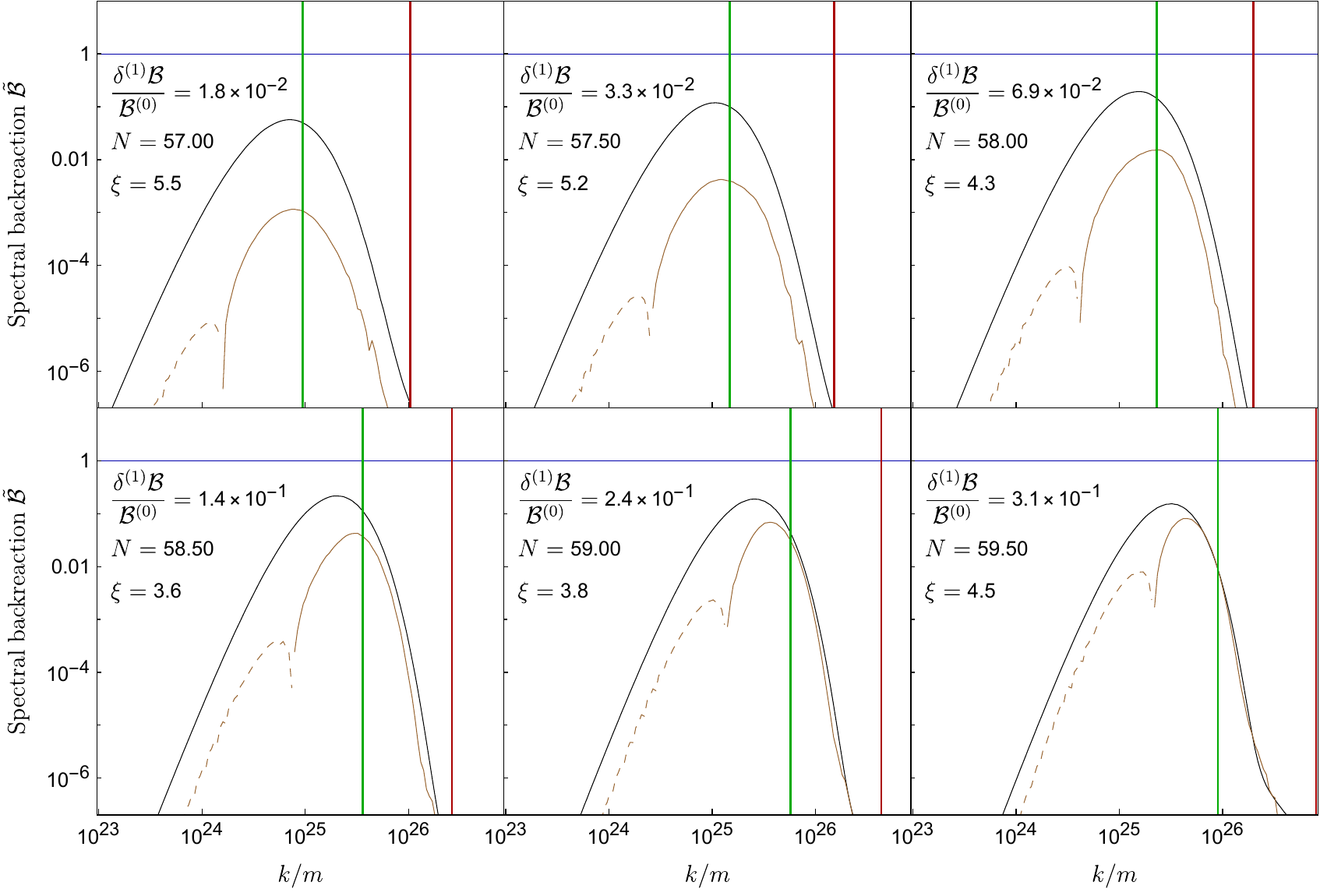}
    \caption{Normalized spectral backreaction for coupling $\beta=25$. Color coding as in Figure~\ref{fig:spec-15}.}
    \label{fig:spec-25}
\end{figure}

%
The series of Figs.~\ref{fig:spec-18} to~\ref{fig:spec-25} displays the evolution of the normalized spectral backreaction for the three intermediate couplings $\beta=\left\{18,20,25\right\}$ studied in this work, all of which are qualitatively similar to the case of the lowest coupling $\beta=15$ shown in the main text.
We note in particular that Fig.~\ref{fig:spec-25} supports the conclusion of an brief period of relatively strong backreaction which remains under perturbative control at $N\simeq58-59.5$.

\clearpage

\bibliographystyle{JHEP}
\bibliography{biblio}

\end{document}